\documentclass[sigconf, nonacm]{acmart}

\usepackage{sparklines}
\usepackage[export]{adjustbox}
\usepackage{subfig}
\usepackage{xspace}
\usepackage{xcolor}
\usepackage{soul}
\usepackage{marginnote}
\usepackage{mdframed}
\usepackage{enumitem}
\newcommand{\ie}{i.e.,\ }
\newcommand{\eg}{e.g.,\ }

\newcommand{\tinyskip}{\vspace{3pt}}
\newcommand{\mypar}[1]{\tinyskip\noindent\textbf{#1.}\xspace}
\newcommand{\myparnoperiod}[1]{\tinyskip\noindent\textbf{#1}}

\newcommand{\F}{\mbox{Fig.\hspace{0.25em}}}

\newenvironment{myitemize}{%
\begin{itemize}[leftmargin=1em, itemsep=.1em, parsep=.1em, topsep=.1em,
    partopsep=.1em]}
{\end{itemize}}

\newenvironment{myenumerate}{%
\begin{enumerate}[leftmargin=1em, itemsep=.1em, parsep=.1em, topsep=.1em,
    partopsep=.1em]}
{\end{enumerate}}

\newenvironment{structure*}{\color{blue}\begin{myenumerate}}{\end{myenumerate}}

\sethlcolor{yellow}
\newcommand{\update}[3][0em]{#3}
\newcommand{\updateL}[3][0em]{#3}

\begin{document}

\title{Data Station: Delegated, Trustworthy, and Auditable Computation to Enable Data-Sharing Consortia with a Data Escrow}

%%
%% The "author" command and its associated commands are used to define the
%authors and their affiliations.
\author{Siyuan Xia, Zhiru Zhu, Chris Zhu, Jinjin Zhao, Kyle Chard, Aaron J. Elmore, Ian Foster, Michael Franklin, Sanjay Krishnan, Raul Castro Fernandez}
\affiliation{\vspace{0.1cm}
\institution{The University of Chicago}
}
\email{{stevenxia, zhiru, chz, j2zhao, chard, aelmore, foster, mjfranklin, skr, raulcf}@uchicago.edu}

\begin{abstract}

    Pooling and sharing data increases and distributes its value. But since data cannot be revoked once shared, scenarios that require controlled release of data for regulatory, privacy, and legal reasons default to not sharing. Because selectively controlling what data to release is difficult, the few data-sharing consortia that exist are often built around data-sharing agreements resulting from long and tedious one-off negotiations.
    
    We introduce Data Station, a data escrow designed to enable the formation of data-sharing consortia. Data \emph{owners} share data with the escrow knowing it will not be released without their consent. Data \emph{users} delegate their computation to the escrow. The data escrow relies on delegated computation to execute queries without releasing the data first. Data Station leverages hardware enclaves to generate \emph{trust} among participants, and exploits the centralization of data and computation to generate an audit log.
    
    We evaluate Data Station on machine learning and data-sharing applications while running on an untrusted intermediary. \updateL{R1W1, R1D1}{In addition to important qualitative advantages, we show that Data Station: i)  outperforms federated learning baselines in accuracy and runtime for the machine learning application; ii) is orders of magnitude faster than alternative secure data-sharing frameworks; and iii) introduces small overhead on the critical path.}
    
    % \updateL{R1W1, R1D1}{We show that Data Station outperforms federated learning deployments, achieving the same or higher accuracy in a small portion of the time. We also show that it reduces by orders of magnitude the time to share data securely while maintaining trustworthiness. We emphasize its qualitative advantages and study its overheads, concluding that they are in most cases small compared to application runtime.}
    
\end{abstract}

\maketitle
\let\thepage\relax 
\pagestyle{plain}

% \begingroup\small\noindent\raggedright\textbf{PVLDB Reference Format:}\\
% \vldbauthors. \vldbtitle. PVLDB, \vldbvolume(\vldbissue): \vldbpages, \vldbyear.\\
% \href{https://doi.org/\vldbdoi}{doi:\vldbdoi}
% \endgroup
% \begingroup
% \renewcommand\thefootnote{}\footnote{\noindent
% This work is licensed under the Creative Commons BY-NC-ND 4.0 International License. Visit \url{https://creativecommons.org/licenses/by-nc-nd/4.0/} to view a copy of this license. For any use beyond those covered by this license, obtain permission by emailing \href{mailto:info@vldb.org}{info@vldb.org}. Copyright is held by the owner/author(s). Publication rights licensed to the VLDB Endowment. \\
% \raggedright Proceedings of the VLDB Endowment, Vol. \vldbvolume, No. \vldbissue\ %
% ISSN 2150-8097. \\
% \href{https://doi.org/\vldbdoi}{doi:\vldbdoi} \\
% }\addtocounter{footnote}{-1}\endgroup
% %%% VLDB block end %%%

% %%% do not modify the following VLDB block %%
% %%% VLDB block start %%%
% \ifdefempty{\vldbavailabilityurl}{}{
% \vspace{.3cm}
% \begingroup\small\noindent\raggedright\textbf{PVLDB Availability Tag:}\\
% The source code of this research paper has been made publicly available at \url{\vldbavailabilityurl}.
% \endgroup

\section{Introduction}
\label{sec:introduction}

Pooling and sharing data increases and distributes its value. Organizations that pool their data can build and mutually benefit from more powerful machine learning models~\cite{learning2017collaborative}. Health organizations that share data with each other can improve patient care~\cite{nightingale}. And, researchers who share experimental data can accelerate scientific discovery~\cite{tenopir2011data}. Despite the obvious advantages, few data-sharing consortia form in practice. Many organizations that could benefit from data sharing face regulatory, legal, privacy, incentive, and technical barriers, and thus, can only release selected data in a controlled manner~\cite{dave2020oblivious, amsterdamer2020towards}. Technically speaking, controlling how data is used is difficult, so many beneficial data-sharing consortia never materialize. Those that do are often built around data-sharing agreements resulting from long and tedious one-off negotiations that are inflexible to later changes in how data should be used.

In this paper, we introduce Data Station, an intermediary \textbf{data escrow}, the computational and data management infrastructure designed to enable the formation of data-sharing consortia. Data \emph{owners} share data with the escrow as if it was an extension of their own infrastructure, \ie it can be guaranteed that their data will remain confidential and that no one will access it (or any derived insights) without their explicit permission. Data \emph{users} who want to extract insights from data \emph{delegate} their computation to the escrow, and that computation will be executed only if permitted by the data owners. The escrow ensures that all data is protected, makes few assumptions on the threat model, and thus, allows owners and users to \emph{trust} it. Finally, because many sharing scenarios involve regulatory and compliance requirements, all computation that takes place on the platform must be transparent so third-party auditors and compliance officers can \emph{audit} the consortia.

\subsection{Data Sharing Scenarios}
We offer stylized scenarios based on examples of real applications to illustrate the opportunities of enabling data-sharing consortia.

\subsubsection{Data Sharing Within Organizations: Team Consortia}
When analysts set out to solve a data task, such as building a machine learning model, or extracting the results of a query to complete a report, they have to find relevant data among a myriad of data sources within their organization~\cite{fernandez2018aurum}. Unfortunately, many of these data sources are siloed and are managed by individual data owners whose responsibility is to control who accesses the data. This introduces an important challenge. Analysts do not necessarily know whether a dataset is useful for their task before seeing it, so they must work with owners in time-consuming one-off negotiations to understand the dataset and negotiate access. Furthermore, even after securing access to a dataset, analysts may find that it is incomplete or poorly described, or that access to additional datasets is necessary, producing a back and forth cycle that increases the time to insights. In short, even within an organization, analysts must invest a significant upfront cost to determine if data is useful. Owners must resort to conversations with the analysts to understand whether data access should be granted.

What would help in this scenario is a platform that permits evaluation of the analysts' tasks on the owners' data without owners having to release the data first. A data escrow would enable owners and users to address the problem while also allowing compliance officers to audit how employees are using data assets and rapidly detect misuse, \eg training ML models over sensitive attributes.

\subsubsection{Data Sharing Across Organizations: Organization Consortia}

Many organizations would mutually benefit from pooling their data to train better machine learning models, but they are wary of sharing their raw datasets. For example, chemical engineering organizations may be willing to pool expensive-to-obtain simulation data to train more powerful models and find better materials~\cite{clifton2004privacy}, but they do not want other organizations to see the data they possess. They would be comfortable sharing inferences over a model trained on everybody's data, but there is no easy solution to orchestrate such data-sharing consortia that does not leak information about their individual datasets. Despite the benefits of sharing, the consequence of the above risks is that organizations do not share data and the value remains untapped.

What is needed is a platform that combines each participant's data without releasing it to anyone, trains a model, and selectively allows participants to access model inferences. Such a platform will help organizations pool their data and unleash its value.

\subsection{The Data Station System}

The main contribution is Data Station, a new data escrow system that enables delegated, trustworthy, and auditable computation.

\mypar{Delegated Computation} Today, data access and processing are intertwined. To run computation, one has to access the data first. However, as data sharing is constrained by the barriers described in the examples, no computation takes place and no value is extracted. Data Station acts as an escrow to whom data owners send their data and data users send their computation. Delegating computation to Data Station means the Station can promise to protect data, \ie no data or derived data products will be released without the data owners' explicit consent. Further, this model stops users from paying upfront costs to access data and allows them to concentrate on gaining access to their query results instead.

\mypar{Trustworthy Computation} The introduction of an escrow enables data sharing as long as both owners and users trust it to keep their data secured. Users must trust that the escrow runs their computation securely and does not leak it to other participants. Owners must trust the escrow to honor their access preferences. Eliciting trust requires different mechanisms that depend on the threat model. Data Station implements a \emph{full-trust} mode: useful in situations such as when employees perform data discovery within their own organization, that runs the intermediary; and \emph{near-zero-trust} mode: useful in cases such as when independent organizations want to pool their data using a third-party intermediary. To implement these mechanisms, Data Station leverages secure hardware enclave technology~\cite{kaplan2016amd} and cryptographic techniques. \update{}{Unlike confidential computing approaches}~\cite{antonopoulos2020azure, arnautov2016scone, haven, lee2020keystone} \update{R2D1}{geared towards letting a user run computation on their own data but on a third-party infrastructure, e.g., a cloud vendor, Data Station is designed to run computation on data from multiple parties.}

\mypar{Auditable Computation} Many challenging data-sharing scenarios are regulated and subject to compliance and audit rules~\cite{edemekong2018health}. In these cases, even if owners and users trust each other, the computation they perform on data must be transparent to third parties, such as compliance officers and auditors. Data Station exploits the centralization of data and compute in the platform to record all computation in a tamper-proof immutable log. Every attempt to access data, every running task, as well as the data access preferences of data owners are stored in the log. This log lets authorized auditors verify that the tasks Data Station runs follow the compliance rules and regulations that govern the data.

\mypar{Contribution and Evaluation Results} \update{R1W1, R1D1}{To the best of our knowledge, Data Station is the first data escrow system designed to enable data-sharing. In the evaluation, we show two sharing applications enabled by Data Station. First, we show that compared to a federated learning deployment, Data Station achieves much higher accuracy in a small portion of the time for several learning tasks. Second, we show how Data Station has up to two orders of magnitude lower overhead than alternative technologies that support end-to-end encryption, thus enabling a wider range of applications. Finally, we emphasize the qualitative advantages of Data Station and conduct a thorough evaluation of its overheads.}

\mypar{Focus of this Paper} Successfully forming data-sharing consortia requires consideration of aspects such as privacy constraints, regulations, legal data-sharing agreements, incentives among participants, and more. All these issues are important, but they matter only if there is a technical solution to share data in the first place. Data Station is designed to tackle the technical challenge. 

The rest of the paper is organized as follows. Section~\ref{sec:primitives} gives an overview of Data Station. Section~\ref{sec:delegatedverifiable} explains how Data Station achieves the goals of delegated and auditable computation. Section~\ref{sec:trust} explains how to support the \emph{near-zero-trust} mode.  Section~\ref{sec:executionenvironment} introduces the design and implementation of an execution environment for Data Station. Section~\ref{sec:evaluation} presents evaluation results, Section~\ref{sec:relatedwork} the related work, and Section~\ref{sec:conclusions} the conclusions.

\section{Data Station Overview}
\label{sec:primitives}

We present the abstractions used by Data Station in Section~\ref{subsec:abstractions}, the sharing lifecycle in Section~\ref{subsec:sharing} and the computation lifecycle in Section~\ref{subsec:computation}. Then, we state the promise Data Station makes in Section~\ref{subsec:promise}, and overview its architecture in Section~\ref{subsec:architecture}.

\subsection{Agents, Data Elements, Functions}
\label{subsec:abstractions}

An agent $a_i \in \mathcal{A}$ is any entity that interacts with Data Station. There are three types of agents. \emph{Owners} control access to data assets they own. It may be beneficial to share these data assets with other agents, so owners will be willing to \emph{register} them with Data Station. Registering a data asset with Data Station copies the data from the owners' infrastructure to Data Station's infrastructure. \emph{Users} are agents who want to run computation on data that is registered with Data Station. Finally, \emph{operators} are neither owners or users, but they want to understand what computation is running on what data inside Data Station. Auditors, compliance officers, and other kinds of regulators may play the role of operators.

Any registered data asset is represented in Data Station as a \emph{data element} (DE), $d_i$. DEs include data of different types and granularities, such as relations, databases, files, images, and more. 

All computation in Data Station is represented via \emph{functions}, $f \in \mathcal{F}$. 
\update{R1W2, R2D4}{Data Station provides some basic functions, but most functions are provided by owners and users planning to form a sharing consortia, e.g., a coalition of chemical engineers may provide specialized indexes and search functionality for molecular data.}
% \steven{or in a machine learning consortia, in which a few parties each holding a subset of the training images or a subset of the training data of the same schema, in which case they can easily provide a function to run on the combined data.}
\update{}{These functions are exposed in Data Station via $f \in \mathcal{F}$}. Functions take input parameters and DEs as input and produce other DEs and optionally other side effects, such as logs and temporal files, e.g., a \textsf{train} function takes input DEs as training data and produces a model. \update{R1W2, R2D4}{No function side effect is visible to users. It follows that functions should run end-to-end.} \update{}{Providing a suite of functions to prepare and integrate data using the escrow is possible, but that requires solving additional challenges out of the scope of this paper.}

\subsection{Policies and Sharing Modes}
\label{subsec:sharing}

Owners register DEs with the Station with the intention of eventually letting some computation run on that data. Owners fully control for what purposes Data Station accesses the data they register via \emph{policies} and \emph{sharing modes}.

\mypar{Policies} A policy is a triple, $a_i, f_i, d_i$ that indicates that agent $a_i$ can run function $f_i$ on DE $d_i$. In contrast to access control policies that broker low-level operations to files, such as read, write, and execution permissions in the context of MAC in Unix-based systems, a policy in Data Station indicates what functions can run on what DEs, so it is a higher-level description.

\mypar{Sharing Modes} There are three \emph{sharing modes} that indicate what types of data access are available: \emph{sealed}, \emph{enclave}, \emph{open}.  In the \emph{sealed} mode, a registered DE cannot be used by any function or by Data Station unless there is an explicit policy permitting such access. When there are no policies, owners who register a DE in \emph{sealed} mode can think of Data Station as a mere extension of their own infrastructure, because no computation can take place and the existence of the DE is not disclosed to anyone.

In \emph{enclave} mode, Data Station can run functions on the DE, but no output will be released without explicit consent from the owner, \ie without an explicit policy permitting the release of the results. This mode permits Data Station to perform tasks such as index creation, profiling, training models, and more, while guaranteeing that no information is released to anyone.

Finally, we say a DE, $d_j$ is in \emph{open} mode for a given agent $a_i$ when there is a policy that includes $a_i$ and $d_j$.

\mypar{Lifecycle} A data owner may initially register a DE in \emph{sealed} mode, but owners register DEs with the intention of eventually allowing users to benefit from their existence. Owners may keep DEs in \emph{sealed} mode and write policies to describe with fine-granularity who can run what computation on their data. This is useful, \eg when allowing a third-party to test a piece of software on their data, or for secure data exchange via an intermediary. In other scenarios, owners may set the DEs to \emph{enclave} mode, letting Data Station run computation on them while keeping results private. This is useful to, \eg build indexes that permit users to discover relevant DEs but ensuring the actual data is never released without explicit consent from the owners. Ultimately, for a DE or derived data product to leave the Station, the data owner must have written a policy explicitly, so that the DE is in open mode.

\subsection{Computation Lifecycle}
\label{subsec:computation}

Users invoke functions pre-registered in the Station, $f_i \in \mathcal{F}$. A function invocation triggers the creation of \emph{intents}, which are triples defined analogously to policies $a_i, f_i, d_i$ and that indicate the intention of agent $a_i$ to execute function $f_i$ on DE $d_i$. Intents are never created over \emph{sealed} DEs unless there is a policy that permits the execution of that function. \emph{Sealed} DEs are invisible to functions.

We differentiate between two broad classes of functions, \emph{data-blind} and \emph{data-aware} functions. The first kind does not require knowledge of any DE in Data Station. For example, search and query-by-example functions take input from a user who does not need to know about any DE. In contrast, \emph{data-aware} functions take as a parameter a set of DEs. For example, a copy/download function needs to indicate what DE to download. It follows that no user can call data-aware functions on (effectively invisible) \emph{sealed} DEs.

% From the definition of \emph{sharing modes} it follows that no user can call data-aware functions on \emph{sealed} DEs because those are to all practical effects invisible, unless an explicit policy is written. Conversely, owners who want to advertise the DEs they registered while still controlling access to those DEs, may choose to set the DEs to \emph{enclave} mode and then allow a function to create an index of derived data products, such as metadata that describes those DEs. Then, users can consult the metadata but they still require a policy to access the underlying DEs.

\mypar{Derived Data Products} When a function runs on a DE and produces an output, we call this output a \emph{derived data product}. A derived data product is another DE that resides inside Data Station. Hence, DEs can be uploaded by their owners, or produced by functions. Derived data products are a key ingredient of delegated computation, as they are the results for which users come to Data Station. One key challenge Data Station must solve is to apply policies created on DEs registered by owners to derived data products, even when these may have been derived from DEs owned by different owners.

\subsection{Data Station's Promise and Trust Modes}
\label{subsec:promise}

Data Station promises owners that only DEs, including derived data products, that are in \emph{open} mode ever leave Data Station. Furthermore, it promises owners that any activity involving DEs they own is recorded and visible to them on-demand. 

Maintaining the guarantee requires different protocols, algorithms, and even infrastructure depending on the threat model. For example, when Data Station runs inside an organization to enable their employees to discover data assets owned by other teams, a reasonable threat model may be that the infrastructure where Data Station is deployed is non-adversarial, that DEs will be kept in their original \emph{sharing mode}, and that the implementation follows the promise as specified. In contrast, when Data Station runs on third-party infrastructure, it must ensure the promise is kept in a more challenging threat model. Data Station operates on different modes to guarantee the promise under different threat models. Agents consider Data Station \emph{trustworthy} when it keeps the promise under their target threat model.

\subsection{Architecture Overview}
\label{subsec:architecture}

\begin{figure}
    \centering
    \includegraphics[width=\columnwidth]{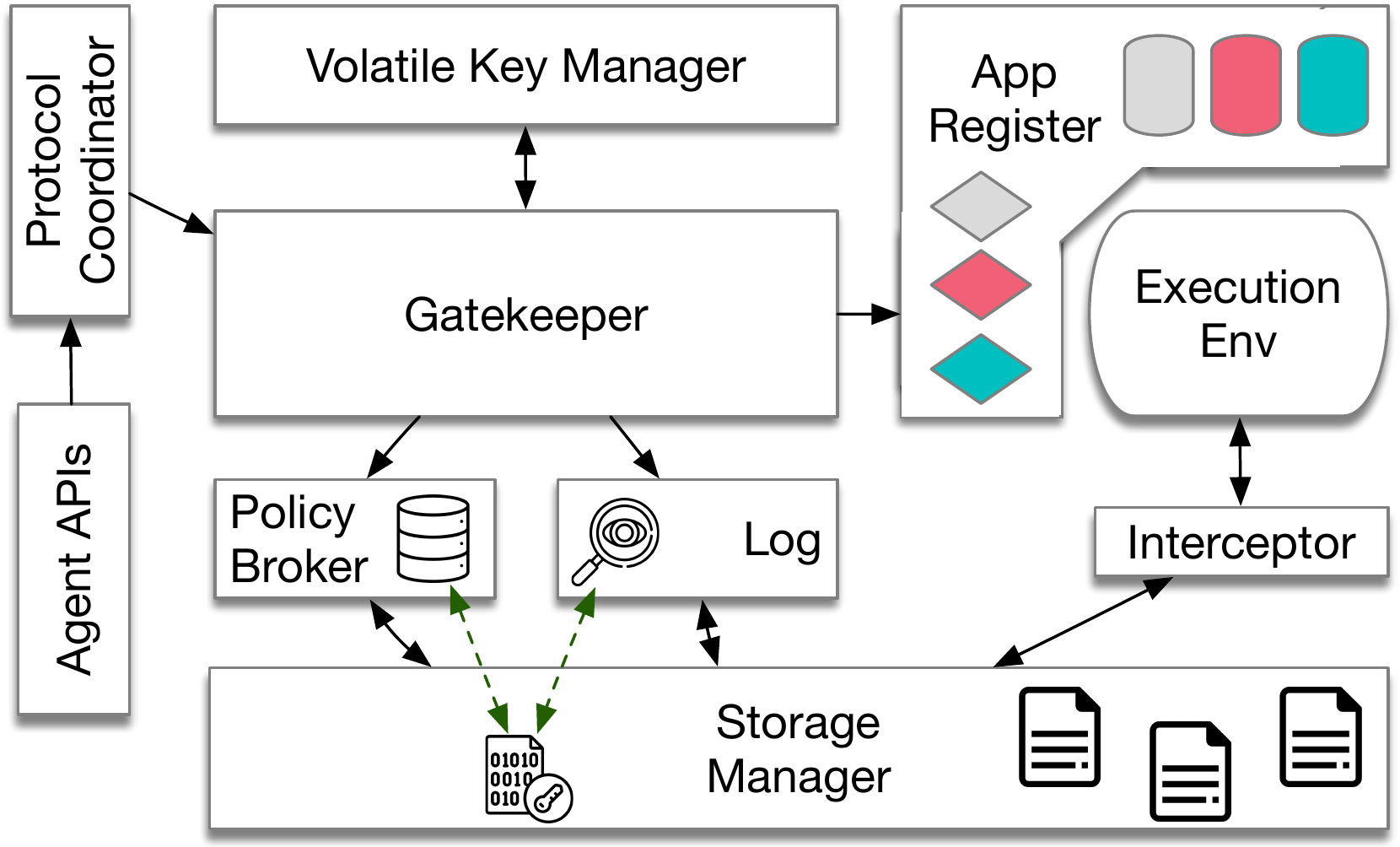}
    \caption{High-level overview of Data Station architecture}
    \label{fig:ds-architecture}
\end{figure}

An overview of Data Station architecture is shown in \F\ref{fig:ds-architecture}. Existing applications can be registered with Data Station via the \textsf{App Register} component that tells the \textsf{Gatekeeper} what functions are available for execution. Once registered, data users can invoke those functions using various interfaces that use the \textsf{Agent APIs} component. All function invocations are brokered by the \textsf{Gatekeeper}, which checks with the \textsf{Policy Broker} if the invocation can proceed, according to the policies present for the data involved and the respective sharing mode. Such policies are written by data owners using the \textsf{Agent APIs}. If the execution can proceed, the \textsf{Gatekeeper} instantiates a task to execute that application, all using the \textsf{execution environment}, where the application runs in isolation. All accesses to the \textsf{Storage Manager}---whether to access existing data or to store derived data products or any other temporary data---is mediated by an \textsf{Interceptor} component. Finally, to support operation under different threat models, Data Station uses a \textsf{Volatile Key Manager} and a \textsf{Protocol Coordinator} that we describe later.

\section{Delegated, Auditable Computation}
\label{sec:delegatedverifiable}

We introduce the main components of Data Station that permit delegated and auditable computation. In this section, we assume both owners and users trust the infrastructure, the function implementation, and the administrators who maintain the platform. We relax these assumptions in Section~\ref{sec:trust}.

\mypar{Goal} The primary goal of Data Station is to run functions invoked by users on data supplied by owners in a way that satisfies owners' sharing preferences, \ie keeping Data Station's promise.

\subsection{Gatekeeper In Depth}
\label{subsec:gk}

After a function invocation, Data Station identifies what DEs are accessible to the function, including derived data products. The \textsf{Gatekeeper} acts as a single-point-of-entry for all function invocations. Thus, it orchestrates all necessary steps to serve function calls while maintaining Data Station's promise. Users can only invoke functions exposed by the \textsf{Gatekeeper}. Functions are registered with the \textsf{Gatekeeper} through an application registration process (Section~\ref{sec:executionenvironment}). Deciding what DEs are available to a function involves a different process for \emph{data-aware} and \emph{data-blind} functions. 

\mypar{Brokering Data-aware Functions} Data-aware functions request access to specific DEs via paths, names, or other identifiers. Ahead of executing the function, the \textsf{Gatekeeper} creates the corresponding intents by using the calling agent id, $a_i$, the function being invoked, $f_j$, and the DE being accessed, $d_k$. The \textsf{Gatekeeper} creates an intent per DE involved. Then, it uses the  \textsf{Policy Broker}  to determine whether the intents have matching policies. The  \textsf{Policy Broker}  component is backed up by a database that contains all agents, functions, DEs, and policies ever registered in the system. Then, given an intent, $(a_i, f_j, d_k)$, the  \textsf{Policy Broker}  creates a query that checks whether there is a policy that permits $a_i$ to execute $f_j$ on $d_k$. For efficiency, whenever more than one intent is created, these are all represented in a single query. Finally, if there are matching policies, the \textsf{Gatekeeper} permits the function invocation on that DE. Otherwise, it blocks execution.

\mypar{Brokering Data-blind Functions} \emph{Data-blind} functions do not include a list of DEs the function needs to access, unlike \emph{data-aware} functions. The \textsf{Gatekeeper} determines what DEs are accessible to the function and agent by combining: i) the set of DEs with a matching policy; ii) the set of DEs in \emph{enclave} mode. Both sets are retrieved from the \textsf{Policy Broker}. The first is by querying DEs that match a predicate containing the calling agent, $a_i$, and function, $f_j$. The second is by requesting DEs stored in \emph{enclave} mode.

\mypar{Function Execution} After determining what DEs are visible to a function, the \textsf{Gatekeeper} must enforce the function only accesses those DEs. This is achieved by creating a ``jailed'' (\ie in the Unix chroot~\cite{kamp2000jails} sense) execution environment that isolates the function from all available resources except for those it is given explicit permission. We explain the design of the execution environment in Section~\ref{sec:executionenvironment}. For now, it suffices to know that such execution environment also captures: i) the concrete set of DEs the function \emph{actually} accesses; and ii) the DEs it produces. First, note that a function may be given access to more DEs than it actually needs access to; for example, a function that builds a spatio-temporal index only needs access to DEs with spatial and temporal attributes. Second, a function will produce and return results. The \textsf{Gatekeeper} must manage the results returned by a function invocation.

\mypar{Staging Zone and Result Delivery} The \textsf{Gatekeeper} must decide whether it can forward the results of a function invocation to the calling agent or not. If the function invoked is \emph{data-aware}, then the function did execute because there is a matching policy for the specific DE indicated. If there is a matching policy, it means that the DE's owner indicated that $a_i$ can execute $f_j$, which implies $a_i$ can access the results. Thus, in this case the \textsf{Gatekeeper} will return the results to the calling agent. If the function invoked is \emph{data-blind}, the function may have executed over both DEs for which there was a policy available, and \emph{enclave} DEs for which there was no policy. When the results depend on only DEs for which there was a matching policy, the \textsf{Gatekeeper} can return them to the calling agent. When they also depend on \emph{enclave} DEs, the results cannot be returned without previously obtaining permission from the \emph{enclave} DE owner. The \textsf{Gatekeeper} determines whether the results depend on only DEs with matching policies or not by retrieving the list of actually accessed DEs from the execution environment, as explained above. Any results that cannot be delivered to $a_i$ directly are stored in the \emph{staging zone}, where they exist as \emph{enclave} DEs, awaiting a matching policy that permits their release.

\mypar{Result Granularity and Brokering Access} When an \emph{enclave} DE, $d_k$, is part of a function's results, the \textsf{Gatekeeper} seeks permission from the data owner to release $d_k$ to $a_i$. The owner originally set the DE in \emph{enclave} mode with the intention of permitting Data Station to perform some computation, but without establishing what functions could be invoked and by whom. After the DE is part of the results of a function invocation, the \textsf{Gatekeeper} provides precise information on both who accessed the DE, $a_i$ and via what function, $f_j$. The owner then may decide to grant access to the calling agent, which amounts to creating the policy $(a_i, f_j, d_k)$. Denying access leaves the list of policies unmodified and the DE removed from the staging zone. Finally, if the list of \emph{enclave} DEs is large, brokering access, which requires involving data owners may become time-consuming. Data Station gives the calling agent the option to call the functions over only the collection of DEs for which a policy is available. This option allows calling agents who want to see the result immediately after it becomes available.

\mypar{Provenance and Granularity} In acting as the single-point-of-entry for all functions, and observing both the input (accessible DEs), the list of actually accessed DEs obtained from the execution environment, and the function output, the \textsf{Gatekeeper} learns the provenance of each derived data product. The provenance is used primarily to understand if the results are accessible by $a_i$, as explained above. The provenance granularity depends on how the applications that register functions are built. For example, a \emph{native} application built using Data Station APIs to access DEs can communicate precisely what DEs were used to produce the different outputs, and how. Data Station permits the execution of unmodified applications. We concentrate on the latter. Here, when the function's output DEs are a subset of the input DEs, the \textsf{Gatekeeper} checks for matching policies. For example, a search function that returns DEs that contain a keyword. When functions return results that are a combination of input DEs, then these newly created DEs, by definition, have no matching policies. For example, when a function trains a machine learning model off training datasets provided by different owners. The \textsf{Gatekeeper} relies on an additional mechanism to aid in handling these latter common cases.

\subsection{Derived Data and Dependency Graph}
\label{subsec:policybroker}

When an intent refers to a DE uploaded by an owner, the \textsf{Gatekeeper} checks if there is a matching policy. But in many scenarios, functions require accessing DEs that are derived from other DEs~\cite{amsterdamer2020towards}. For example, a search function may use an index, which may in turn be created by another function from a set of accessible DEs. The index is an intermediate DE. Because of that, no owner writes directly a policy for the index. When functions run on intermediate DEs, Data Station needs a mechanism to determine access to DEs and to facilitate the task of data owners when writing policies.

\mypar{Function Dependency Graph} The dependency graph is a directed acyclic graph (DAG). Nodes represent functions, and directed edges represent dependencies between these functions in the form of DEs.  For example, a \emph{search} function depends on an \emph{index} function, that itself creates the index from DEs. The dependency graph is created during the application registration process (Section~\ref{sec:executionenvironment}). Owners understand the dependency between functions. When they write a policy to permit a function to execute on DEs they own, they are implicitly permitting any children of the function to access that DE as well. Hence, if they write a policy for the \emph{search} function, they are also allowing \emph{index} to access their DEs.

\mypar{Policy Matching with Dependency Graph} To leverage the function dependency graph, during the policy matching process, the \textsf{Gatekeeper} must check whether there is an existing policy for the function, or any descendant. Without the dependency graph, if there is no matching policy, the \textsf{Gatekeeper} would only grant execution access to the function if the DE is in \emph{enclave} mode. And this would later require brokering access to the function results, introducing unnecessary delays. Instead, the dependency graph permits skipping unnecessary handling of policies and vastly simplifies the task both for users, owners, and the \textsf{Gatekeeper}.

\subsection{Auditable Computation: The Log}
\label{subsec:thelog}

If Data Station keeps its promise, then owners are guaranteed that their DEs are accessed only in the way the policies they wrote permit. There are scenarios where auditors, compliance officers, and other third parties need access to the inner doings of Data Station. Because Data Station centralizes computation and data, they capture the provenance of every function invocation. This can be stored in a log and offered to operators as a source-of-truth record of what Data Station has done.

To be useful to operators, the auditability log must record \emph{every} intent, policy match check, and DE result delivery that Data Station performs. It must permit data owners consulting the information that concerns DEs they own and it must permit data operators accessing this information when the participating agents have agreed to such arrangement. In other words: no computation can occur in Data Station without it being recorded in the log.

The \textsf{Gatekeeper} is the perfect candidate to manage this log because it already acts as a central actor checking every function invocation that aims to access DEs. Thus, the log, architecturally, resides inside the \textsf{Gatekeeper}. All writes to this log originate in the \textsf{Gatekeeper}, and all reads to the log go through the \textsf{Gatekeeper}, as with any other function invocation.

The log is a DE whose owner is Data Station. When a new owner joins Data Station, Data Station creates policies that permit owners to access the log and consult any activity that pertains to DEs they own. The owners can then inspect the log by invoking the corresponding function.

% Then, when these owners want to inspect the log, they invoke the appropriate function, the \textsf{Gatekeeper} creates an intent, and finds the intent matches an existing policy that in turn opens access to entries concerning DEs they own.

\mypar{Log Entries} The log resides on disk and consists of a sequence of entries. Each entry consists of an agent id that corresponds to the caller that triggered the creation of the entry, and a payload. The payload can be of different types. It can indicate an intent was created, an intent-policy match, and a mismatch. And it also records what DEs are allowed outside the Station, including derived data products. Note that it is possible to trace what DEs contributed to any intermediate DE because the \textsf{Gatekeeper} keeps the provenance, which is itself materialized in this log. All history concerning DEs is stored in the log and this is the source of auditability.

\mypar{Opening Access to Third Parties via Contracts} By default, the only agents who can inspect the log are owners. And they can only consult entries related to DEs they own. To permit access to third parties, all participating agents in Data Station must produce a contract. With a contract, Data Station creates a policy for the operator agent, who can then consult the log. A contract is a policy $(a_o, r, l)$, where $a_o$ is the id of the operator, $r$ the read function on the log, and $l$ the DE that refers to the log. A contract is different than a policy in that it must be signed by every participant. Without a signature per participant, Data Station does not create the policy, and the operator cannot consult the log. Data Station relies on public key cryptography to permit agents to sign contracts; these primitives are introduced in the next Section.

\section{Trustworthy Computation}
\label{sec:trust}

We introduce mechanisms used to protect Data Station's promise when running on untrusted infrastructure. We say Data Station runs in \emph{near-zero-trust} mode when these mechanisms (introduced in Section~\ref{subsec:principles}) are activated. We present the encryption protocols used by Data Station in Section~\ref{subsec:protocols} and conclude by explaining how we deal with the log and the databases (Sections~\ref{subsec:logrecovery} and \ref{subsec:dbrecovery}).

\mypar{Threat Model} We assume that a curious operator gains access to Data Station infrastructure and can read disk and memory contents. To keep its promise, Data Station cannot leak any DE to this operator. Furthermore, Data Station cannot leak information from the database (that contains all agents, DEs, and policies) or auditable log. The adversary may gain access to a list of agent ids, but they should not be able to link those ids with any other information. We do not protect against denial-of-service attacks.

\subsection{Near-Zero-Trust Principles}
\label{subsec:principles}

\update{R2D2}{To ensure the confidentiality of every DE in Data Station, data is encrypted end-to-end, from the moment where it leaves the agent's infrastructure, and including while functions access that DE in memory, i.e., during processing. To ensure integrity, every message is signed with the private key of the agent from where it originates. To bootstrap trust, Data Station's node proves its identity to agents and attests that it runs the original Data Station software, and not a modified version, thus avoiding backdoor attacks.}

Every shared DE is encrypted with an agent's specific symmetric key. It is then transmitted to Data Station over a secure channel with TLS~\cite{dierks2008transport}. Once in Data Station, it remains encrypted at rest. When a function needs to access the data, Data Station leverages secure hardware enclaves to maintain the data encrypted in memory, \ie for processing, the DE is decrypted into encrypted memory. Functions may need to store intermediate results in the file system, such as temporal files. Because some of these intermediates may leak sensitive information, these intermediates are encrypted transparently to the application: no function can store plain data on disk when Data Station operates on \emph{near-zero-trust} mode. We introduce secure hardware enclaves and the two key properties Data Station uses: encrypted memory and remote attestation.

\subsubsection{A Primer on Secure Hardware Enclaves}

AMD's Secure Encrypted Virtualization (SEV) and Intel's Software Guard eXtensions (SGX) leverage specially-built hardware to isolate virtual machines (node) \cite{kaplan2016amd} and applications \cite{inteldevguide3d}, respectively, within areas called \emph{enclaves} to protect data leakage from even privileged users of the system. These technologies introduce important tradeoffs.

Compared to SEV that encrypts all of a node's working memory, SGX limits the total working memory to 128MB \cite{intelmeasurementsgx}. The upside is that with SGX users only need to trust the application that runs inside the enclave. In contrast, with SEV they must also trust the OS, which is ultimately responsible for ensuring memory pages are encrypted. Another downside of SGX is that applications need to be rewritten, as opposed to SEV, which accepts unmodified software. Since January'22, Intel has deprecated SGX \cite{inteldatasheet}. This, in addition to the increased convenience of having all memory encrypted means we implement Data Station leveraging AMD's SEV.

\mypar{Encrypted Memory} AMD's SEV guarantees confidentiality by encrypting all of the OS's writes to memory. \update{R2D2}{Unauthorized users (including by the hypervisor in cloud contexts) cannot read data in plaintext, i.e., dumping the memory contents of the process (e.g., \texttt{cat /proc/[pid]/maps}) will show a cyphertext}. With the more recent SEV-SNP (Secure Nested Paging) \cite{sev2020strengthening}, the Trusted Computing Base (TCB) (the set of all hardware, firmware, and software that agents need to trust) only has two components: the AMD hardware and firmware, and the operating system image running in the node. All other components are untrusted, including the BIOS, hypervisor, other images (in the case of multi-tenant cloud scenarios), and external PCI devices. In summary, agents need only trust AMD's hardware is correctly implemented and that the OS image implements SEV correctly.

\mypar{Attacks to Enclave Implementations and Limitations} Data Station operating in \emph{near-zero-trust} inherits any vulnerabilities of the underlying SEV-SNP implementation. Solving SEV's implementation specific bugs~\cite{li2021crossline, li2021cipherleaks} is outside the scope of this paper, and AMD actively works to mitigate them at the time of writing. \update{R1W3}{SEV enclaves protect the confidentiality and integrity of data in main memory, but not of data living on external devices, such as disks or GPUs. Data Station offers protection for data on disk as well as its transfer to memory, but it does not support computation on the GPU. Another limitation of enclaves is their reduced performance: we show in the evaluation section that this overhead does not affect application runtime significantly. Finally, every major vendor provides a secure enclave technology, but the specific security guarantees of SEV that make it a good fit for Data Station are, as of May'22, only provided by AMD and available in the Google Cloud.}

% \update{R1W3}{Furthermore, while enclaves (SEV) protect the confidentiality and integrity of data in main memory, they do not protect data living on external devices, such as disk, buses, and GPUs. Thus, the platform implementing SEV must take additional measures to protect data when outside the main memory of the machine. Data Station offers protection for data on disk as well as its transfer to memory, but it does not support computation on the GPU. Other limitations commonly attributed to hardware enclaves are their reduced performance. While they introduce a performance penalty, we show in the evaluation section that this overhead is small and thus does not affect application runtime significantly. Finally, every major vendor provides a secure enclave technology, but the specific characteristics of SEV that makes it a good fit for Data Station is as of May'22 only provided by AMD and available in the Google Cloud.}

\subsubsection{Bootstrapping Trust with Remote Attestation}

To convince agents that the infrastructure running Data Station has SEV enabled and that the software running is indeed Data Station software, the platform uses remote attestation, as provided by AMD's hardware. This permits agents request a report from the node that contains unique identifying information. Furthermore, the agents can trust the node is running a version of Data Station with a correct implementation of the software, including the \textsf{Gatekeeper}, \update{R2D2}{as opposed to an adversarial modified version  that includes a backdoor. If the software restarts, the entire remote attestation process repeats to avoid an attacker swapping software versions.} 

% Assuming agents trust the trust the Data Station's code is as described, for each Data Station component to the SEV node, remote attestation allows an agent to verify that software running on the Data Station is exactly what was uploaded and agreed upon, \ie that the gatekeeper is indeed following the correct protocol.

\subsection{e2e Encryption Protocol and Key Manager}
\label{subsec:protocols}

\update{R1W4, R1D2}{The end-to-end encryption protocol must ensure all DEs are always encrypted, including during processing. At the same time, it must permit agents with a matching policy to access the DEs. Two components are primarily responsible for achieving this goal, the \emph{Protocol Coordinator} that uses mostly standard public-key cryptography, and the \emph{Volatile Key Manager} that is the mechanism used to protect cryptographic keys. We explain both next.}

\mypar{Preliminaries} \update{}{Every agent and Data Station have a public and private key. In addition, every agent has a symmetric key that they share with Data Station. Data is encrypted and signed (to prove identity ) at origin and remains encrypted throughout the lifecycle.} 

\mypar{Protocol Coordinator} \update{}{To process a DE, Data Station first decrypts it in memory by using the symmetric key shared by the DE's owner. Because memory in the enclave is encrypted, the data remains protected from external observers. To transmit a DE (e.g., to an agent with a matching policy) Data Station first decrypts the DE in memory and re-encrypts it with the symmetric key of the receiving agent. Again, the DE remains encrypted at all times because this processing takes place in-memory.}

\mypar{Volatile Key Managemer} \update{}{Data Station possesses two classes of sensitive information: i) the symmetric keys used by agents to encrypt the DEs; ii) its own private key. If any of these keys is compromised, the whole system's guarantees fall apart. To protect these keys, they are stored in a volatile key manager that resides in-memory, hence encrypted. This is a simple way of maintaining the keys secure while Data Station is running. However, if Data Station restarts (e.g., failure, maintenance), the keys will vanish from the key manager, so a strategy to recover them is needed. Some enclave technology, such as Intel's SGX supports \emph{sealing} data, which means encrypting in-memory data to disk with an enclave-specific key. AMD SEV does not support \emph{sealing}~}\cite{sevguide}. \update{}{Hence, Data Station relies on agents resending their encryption keys to recover the Key Manager state when necessary.}

% \mypar{Volatile Key Manager} The Data Station keeps all agent keys, public and symmetric, in an in-memory store. Because the memory of the enclave is encrypted, the data is protected. However, if there is a failure or the Data Station must restart, the keys will vanish. Some enclave technology, such as Intel's SGX support \emph{sealing} data, which means encrypting in-memory data to disk with an enclave-specific key. AMD SEV does not support \emph{sealing}~\cite{sevguide}. Hence, Data Station relies on agents resending their encryption keys to recover the Key Manager state when necessary.

\mypar{Derived Data Products} Data Station is responsible for encrypting derived data products before storing them on disk. Notice that there is no need to use any specific key. Unlike with shared DEs, derived data products exist because owners let Data Station invoke functions on them. It follows Data Station can select what key to use to maintain their encryption on disk. Data Station uses the calling agent's symmetric key to avoid one round of re-encryption if the calling agent is given permission to access the DE.

\subsection{Audit Log Management}
\label{subsec:logrecovery}

If the audit log was stored in plain text, an attacker who gains access to the node would learn what functions were executed on what DEs and by whom, and thus could learn about existing policies. Although this may not be critical in some applications, Data Station protects against this risk by maintaining the log encrypted on disk. The log is a sequence of $\langle a_i, E_{k_u}(l) \rangle$ entries. The agent ID, $a_i$, is in plain text, but it is only meaningful with access to the database.

Only owners can see log entries that involve DEs that they own, so naturally one may think that these entries are encrypted with the owner's key. However, an entry may involve many DEs, \eg functions that require accessing collections of DEs, and encrypting the entry with each DE's owner's key is inefficient when the number of DEs is large. Instead, Data Station encrypts the log entries with the symmetric key, $k_u$, of the data user whose function call triggered these entries. Because only Data Station can access the audit log and each entry is generated by exactly one agent, this solution is more efficient. 

% 

% In the event of failure, or reboot of the Data Station, all keys disappear from the volatile key manager, and consequently, access to the encrypted audit log is lost. To regain access, the Data Station must bootstrap the re-population of the key manager until it contains keys for all agent ids that appear alongside the log entries. Doing so requires access to the database, the other system DE that Data Station needs to manage in \emph{near-zero-trust} mode.

\subsection{Database Management}
\label{subsec:dbrecovery}

An attacker who gains access to the infrastructure where the database is hosted learns all policies, DE locations (but not access, as these are encrypted), and information about any agents registered with the platform. To protect against this, the database remains in-memory, and hence, encrypted with SEV. Then, even when an attacker gains access to the physical machine, the contents of the database are protected. The challenge then is how to deal with failures and reboots of Data Station.

To solve this problem, Data Station uses an encrypted write-ahead-log (EWAL) that is independent of, and external to any WAL used internally by the database system. All updates to the database are first stored encrypted in the EWAL. The entries are encrypted by using the symmetric key of the agent who generated the update. Each entry is stored along with the agent id that caused the update. The assignment of IDs to agents takes place outside the database so these can be incorporated in the EWAL log entries. After restarting, the database can be recovered from the log, as in the traditional recovery protocols of relational databases~\cite{mohan1992aries}. To deal with a growing EWAL, the database can be checkpointed to disk periodically. Ahead of storing the checkpoint on disk, this must also be encrypted in memory (see below). After a restart, decrypting the EWAL entries requires collecting keys from the agents first. After the EWAL is replayed, Data Station is considered recovered. With a recovered database, Data Station can recover, in turn, the audit log.

% A final technical detail is that  -- Otherwise, if the system fails after a new agent is registered but before the database is checkpointed, the corresponding log entries will be locked.

\mypar{Encrypting Database Checkpoint without a Sealing Mechanism} The database checkpoint must be encrypted ahead of being stored on disk. With sealing, Data Station would use the enclave key to encrypt the checkpoint. Without a sealing mechanism, we resort to a different solution. The baseline solution is to select one of the symmetric keys of agents to encrypt the checkpoint. On restart, agents will resend their keys to Data Station, which attempts to decrypt the checkpoint with each newly received key until it succeeds. Because operation cannot be resumed until all agents have resent their keys, this process does not introduce additional delays. To increase reliability, \ie in case the agent whose key encrypted the checkpoint does not reconnect to Data Station, the checkpoint can be encrypted with $m$ agents' keys instead, and then as soon as one key decrypts the database Data Station becomes operational.

\subsection{Other Protections and Limitations}
\label{subsec:limitations}

Potential adversaries that gain control of the infrastructure cannot modify the contents of the database because it is encrypted. Hence, they cannot create policies, agents, or change the sharing mode of existing DEs. Furthermore, these attacks to the integrity of the memory contents will be discovered when \emph{SEV-SNP} becomes available. All interactions with Data Station are mediated via the \textsf{Protocol Coordinator}, who among other things, handles authentication.

Although the EWAL and audit log are encrypted on disk, an attacker could potentially perturb these, \eg by adding random bytes. Data Station is not protected against such denial of service attacks. This limitation of the implementation could be addressed by extensions to support a replication service.

% The database is encrypted $m$ times using a random sample of $m$ agents' keys. On reboot, the Data Station proceeds to recover keys from the agents and attempts to recover the checkpoint that way. The checkpoint is recovered as soon as the Data Station obtains 1 of the $m$ keys used to encrypt the contents. Although this mechanism may seem inefficient, it is necessary to overcome the lack of \emph{sealing} in AMD's SEV and it only needs to run occasionally after a failure.

\section{Execution Environment}
\label{sec:executionenvironment}

% We present the requirements and implementation of an execution environment for Data Station.

% \subsection{Requirements}

\mypar{Requirements} The execution environment comprises all the functionality and components of Data Station that permit the execution of applications. In designing Data Station, we wanted to permit existing applications execute unmodified to ease their deployment. The requirements for the execution environment are:

% \begin{myitemize}

\noindent$\bullet$ Developers \emph{register} existing applications with Data Station without making changes to the application's implementation. They provide a simple \emph{connector} that indicates how to invoke the application functionality.

\noindent$\bullet$ Data Station provides the application with the necessary resources to execute and serve function invocations while guaranteeing the application is isolated from other applications and from data it cannot access.

\noindent$\bullet$ During execution, the DEs accessed by the application must be recorded by the execution environment and sent to the \textsf{Gatekeeper} after the function finishes.

\noindent$\bullet$ In the \emph{near-zero-trust} mode, data is encrypted on disk. The execution environment ensures applications access decrypted data transparently so they do not break during execution.

% \end{myitemize}

% Next, we explain the design and implementation of an execution environment.

\subsection{Design}

\mypar{Function Registration} Application developers \emph{register} functions with the \textsf{Gatekeeper} via a \emph{connector}. The connector contains the functions exposed to Data Station and the dependencies between these functions. This is all the information Data Station needs to build the \emph{function dependency graph}. Upon system initialization, Data Station loads all connectors from developers to register the functions with the gatekeeper. Each function in the connector is in charge of invoking the functionality from the application.

\mypar{Resource Management and Isolation} When the \textsf{Gatekeeper} grants execution permission to a function, the function is instantiated in an isolated process with restricted and controlled access to the system's resources. 
% In particular, it ensures it can only access certain paths on disk.

\mypar{Interceptor Middleware} Data Station starts an \textsf{Interceptor}  upon initialization. The \textsf{Interceptor}  knows all processes started by the \textsf{Gatekeeper} and the locations such processes can access. It intercepts all I/O calls from the function's isolated process to storage. The \textsf{Gatekeeper} and the \textsf{Interceptor}  middleware work in a client-server manner as follows:

% \begin{myitemize}

\noindent$\bullet$ The \textsf{Gatekeeper} passes the list of accessible DEs to the \textsf{Interceptor} through the function's execution environment. The \textsf{Interceptor}  makes sure that only those accessible DEs are visible to the function by filtering out those that are not.

\noindent$\bullet$ The \textsf{Interceptor} records all DEs actually accessed by the function and sends them back to the Gatekeeper, again through the function's execution environment.

\noindent$\bullet$ When operating in \emph{near-zero-trust} mode, data is encrypted on disk, and no function side effects (such as temp files) should be stored in plain text. Whenever a function invokes a read operation, the \textsf{Interceptor} decrypts the data on-demand; analogously, it encrypts writes using the data owner's symmetric key.

% \end{myitemize}

Last, the \textsf{Gatekeeper} obtains from the \textsf{Interceptor} the list of DEs the function actually accessed and proceeds as explained earlier.

% After the function is done, the \textsf{Gatekeeper} asks the \textsf{Interceptor} to send the DEs that this function invocation actually accessed and proceeds as explained earlier.

\subsection{Implementation}

The executor environment operates on a local file system. Extending it to other settings is beyond the scope of this paper.

The \textsf{Gatekeeper} creates isolated processes using Docker containers. Each container effectively creates a jail (in the chroot sense) that limits the resources accessible by functions. The container is created so that all paths the function can access are intercepted by the Interceptor. Furthermore, instantiating functions in Docker containers permit easy management of the resources available.

We implement the \textsf{Interceptor} middleware by using FUSE~\cite{szeredi2010fuse}, with the libfuse userspace library~\cite{pythonfuse} %\footnote{https://github.com/libfuse/python-fuse} 
used to create file systems in userspace. During Data Station's system initialization, the \textsf{Interceptor} mounts Data Station's storage to a specified mountpoint. All subsequent function invocations will access DEs through the mountpoint, so that all I/O operations can be intercepted.

Using FUSE has one key advantage---since the filesystem is created in user space, we do not need to modify the kernel code to intercept the I/O calls and perform additional operations. Avoiding kernel modification is necessary to reduce the size of the trusted computing base. The FUSE filesystem offers a flexible way to interact with Data Station without compromising its security guarantees.

To support concurrent function invocations, the \textsf{Interceptor}  works as a server that accepts requests from the gatekeeper, which starts an isolated process per function invocation. The \textsf{Interceptor}  keeps track of the corresponding DEs accessed by each running function. This is achieved by associating each data access with the identity of the execution environment that attempts to access the data. Similarly, the list of accessible DEs (and the corresponding symmetric keys if operating in near-zero-trust mode) are also associated with the identity of an execution environment. The \textsf{Gatekeeper} and \textsf{Interceptor}  must remain connected to ensure the right process contextual information is shared.

\section{Evaluation}
\label{sec:evaluation}

\begin{figure*}[]
  \centering
  \subfloat{
  \includegraphics[width=0.33\linewidth]{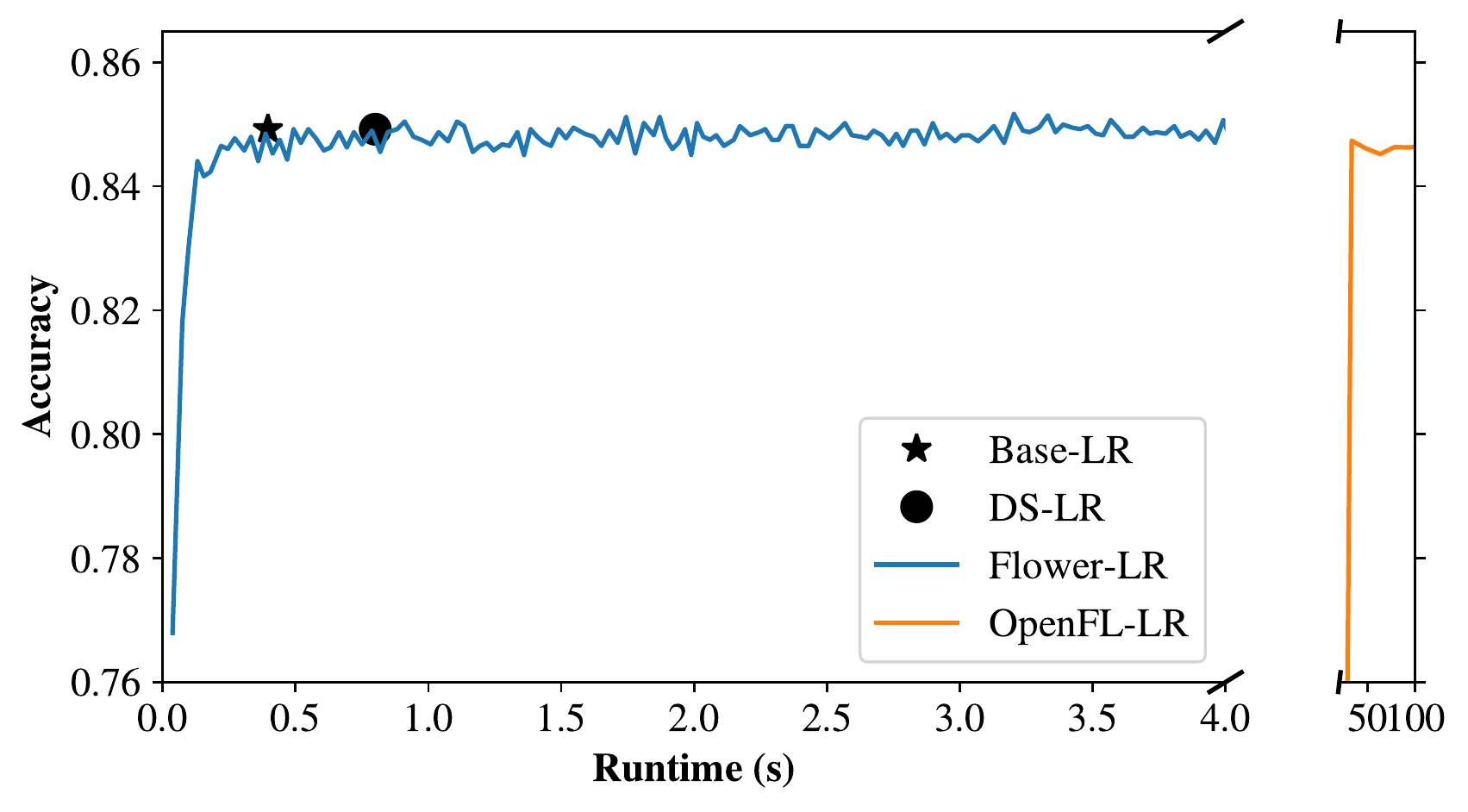}
  \label{fig:ml_income}
  }
  \subfloat{
  \includegraphics[width=0.33\linewidth]{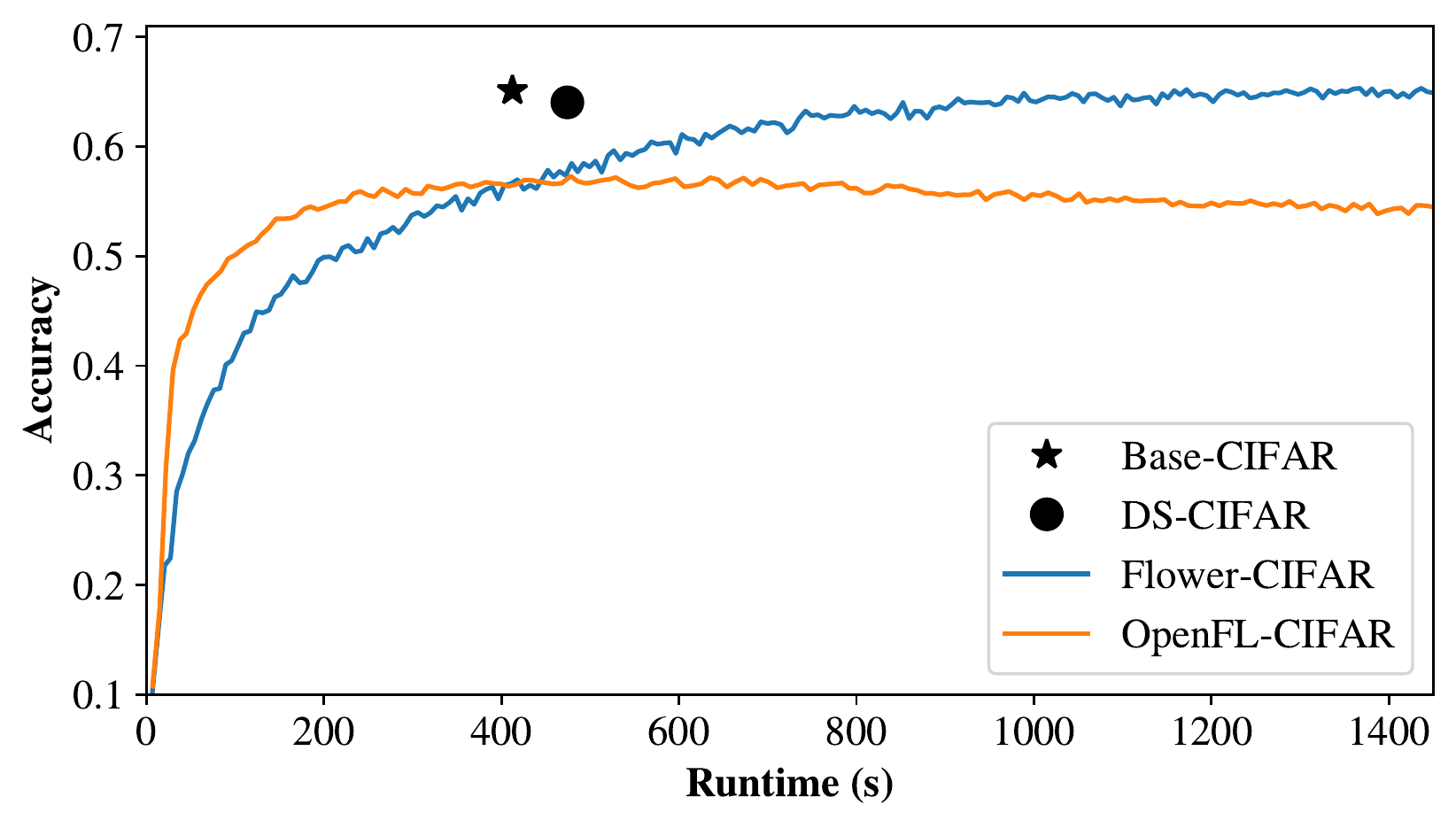}
  \label{fig:ml_cifar}
  }
  \centering
  \subfloat{
  \includegraphics[width=0.33\linewidth]{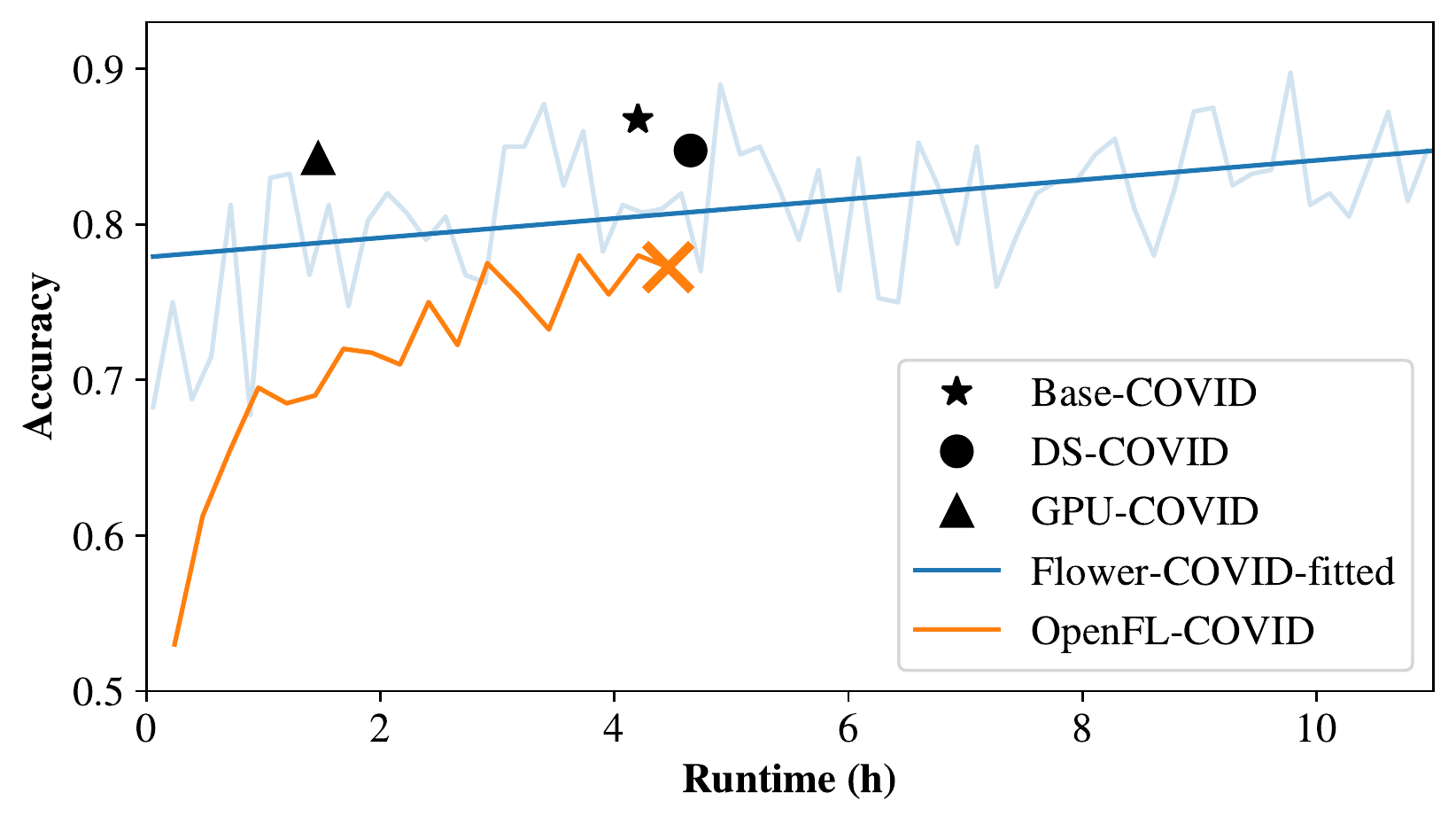}
  \label{fig:ml_covid}
  }
  \centering
\caption{Accuracy and runtime of the four baselines on adult income (left), CIFAR-10 (center), and COVIDx (right).}

\label{fig:rq2_stability}
\end{figure*}

We present the evaluation results to answer two questions:

% \begin{myitemize}

\myparnoperiod{RQ1: What applications does Data Station enable?} We implement two types of applications in Data Station: machine learning applications and a file-sharing application. We find that Data Station achieve much better quantitative results and present a number of invaluable qualitative advantages.

\myparnoperiod{RQ2: Do Data Station's design decisions lead to a practical system?} \update{R2W3}{We study the overheads introduced by the need to provide trustworthiness and the use of SEV. Our aim is to determine whether such overheads pose a runtime bottleneck to applications. We study each component in Data Station as well as the overhead introduced by SEV. We show that overheads are negligible even when running in \emph{near-zero-trust} mode.}

% \update{R2W3}{To achieve trustworthiness, Data Station makes a series of design decisions and relies on SEV. This question explores whether the overheads introduced by the Data Station design lead to a system practical for real applications. Thus, we study the overheads of Data Station and SEV and frame the results in the context of the applications' runtime.} Besides, we conduct a series of experiments and microbenchmarks to 
% explore the overheads introduced by each component. We then contextualize these overheads with respect to the runtime of typical applications to discover that the overheads are negligible, even when running the \emph{near-zero-trust} mode.

% \end{myitemize}

\mypar{Outline} We divide our evaluation into three subsections, the first two address \textbf{RQ1} and present the quantitative and qualitative results for two applications. The last presents the characterization of overheads of Data Station, thus addressing \textbf{RQ2}.

% \mypar{Experimental Setup} We deploy Data Station on Google Compute Engine using a n2d-highmem-8 instance (8 virtual CPUs and 64GB of RAM). We use an OS image with SEV enabled when using Data Station. When running baselines that do not require hardware enclaves, we disable SEV to avoid unnecessary overheads.

\subsection{Machine Learning Consortium}

We consider a scenario where $N=8$ agents want to pool their individual datasets, $D_i$, to train a more powerful machine learning model but without allowing other agents to see their raw data. Only access to inferences on the jointly trained model is permitted. \update{R1D3}{When using Data Station (in \emph{near-zero-trust} mode), each agent registers their data with the platform because they know their data will be protected end-to-end. We also consider a federated learning scenario, where agents' data never leaves their own machine and the model is trained in a distributed manner. The federated learning setting corresponds to the closest setting we can deploy to achieve the desired goal, and we compare federated learning with Data Station from a qualitative perspective later in this section.}

\mypar{Baselines}\update{R1W5, R1D3}{We consider the following baselines:}

% \notera{need to add 1 sentence for flower and one for openfl}

% \begin{myitemize}

\noindent$\bullet$ \update{}{\textsf{Base}. This is a centralized untrusted server training the model. It serves as a reference to compare with other baselines.}

\noindent$\bullet$ \update{}{\textsf{DS}. Data Station running in \emph{near-zero-trust} mode.}

\noindent$\bullet$ \update{}{\textsf{Flower}}~\cite{beutel2020flower}. \update{}{A state-of-the-art federated learning framework implementing the \textsf{FederatedAveraging} aggregation algorithm}~\cite{fedaverage}. \update{}{We sample from 2 clients after each round to accelerate convergence and as recommended in the documentation. We deploy Flower in a distributed setting, with each client/agent accessing its own node.}

\noindent$\bullet$ \update{}{\textsf{OpenFL}}~\cite{reina2021openfl}. \update{}{A state-of-the-art federated learning framework that implements the same aggregation algorithm as Flower but does not sample per epoch. Its support for distributed computing is not mature; we deploy all clients in a single node.} 

% \steven{How is it not robust? Do we need to explain it? I see that this point is discussed in more depth later. Should we move those discussion up here?}

% \end{myitemize}

\update{R1W5, R1D3}{When exploring federated learning baselines we considered others such as PySyft}~\cite{ziller2021pysyft} \update{}{and FATE}~\cite{FATE}. \update{}{We choose Flower and OpenFL because they implement state-of-the-art federated aggregation algorithms and are mature: i) they are actively maintained; ii) their documentation is complete; iii) the examples in their tutorials work.}

% We assume 8 clients, each with private data, want to train a joint ML model without releasing their raw data. Clients trust the federated learning setting because their raw data never leaves their platform. Clients trust Data Station running in \emph{near-zero-trust} mode because their data remains encrypted end-to-end.}

\begin{figure*}[]
  \centering
  \subfloat{
  \includegraphics[width=0.33\linewidth]{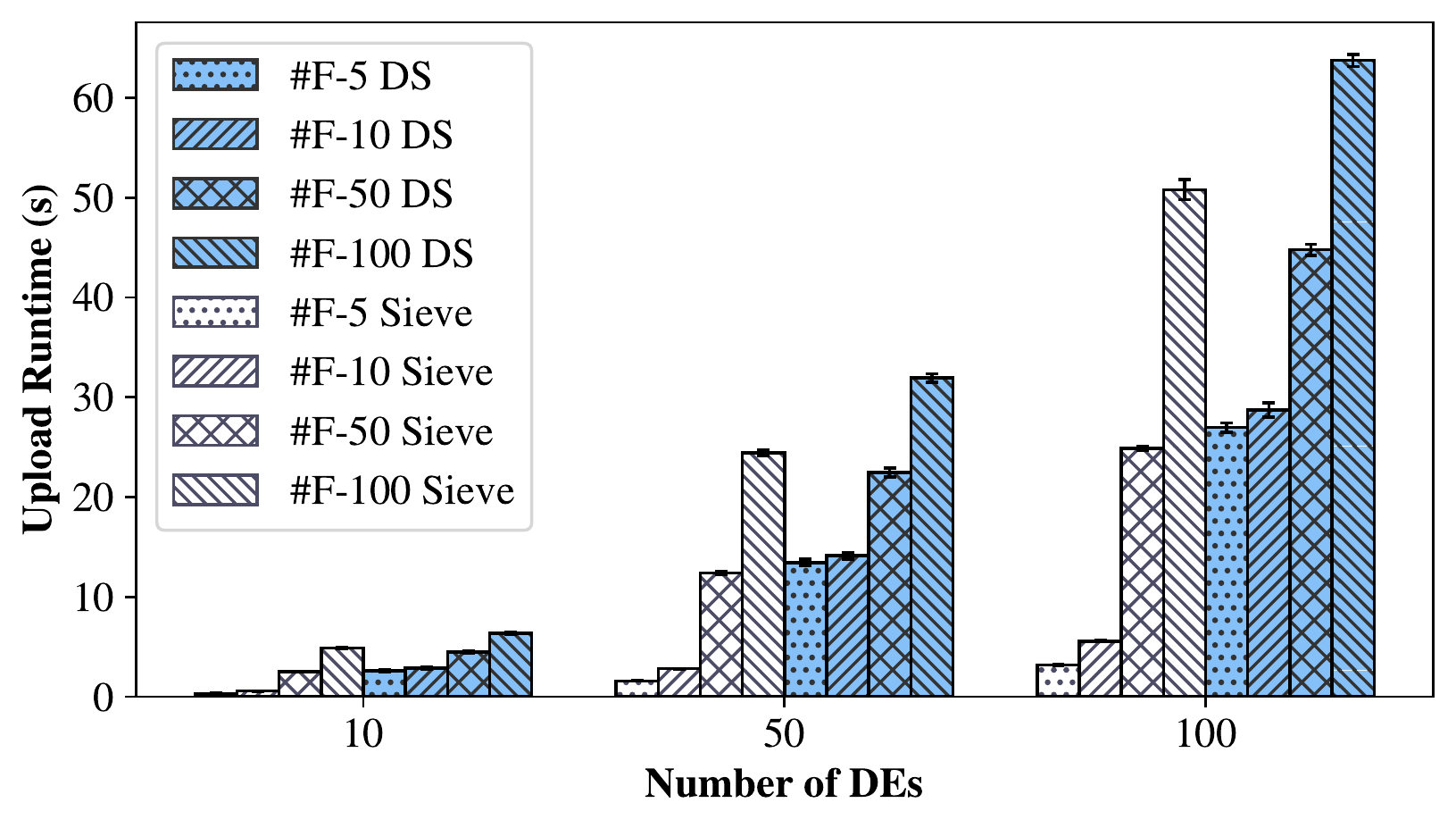}
  \label{fig:ds_sieve_upload}
  }
  \subfloat{
  \includegraphics[width=0.33\linewidth]{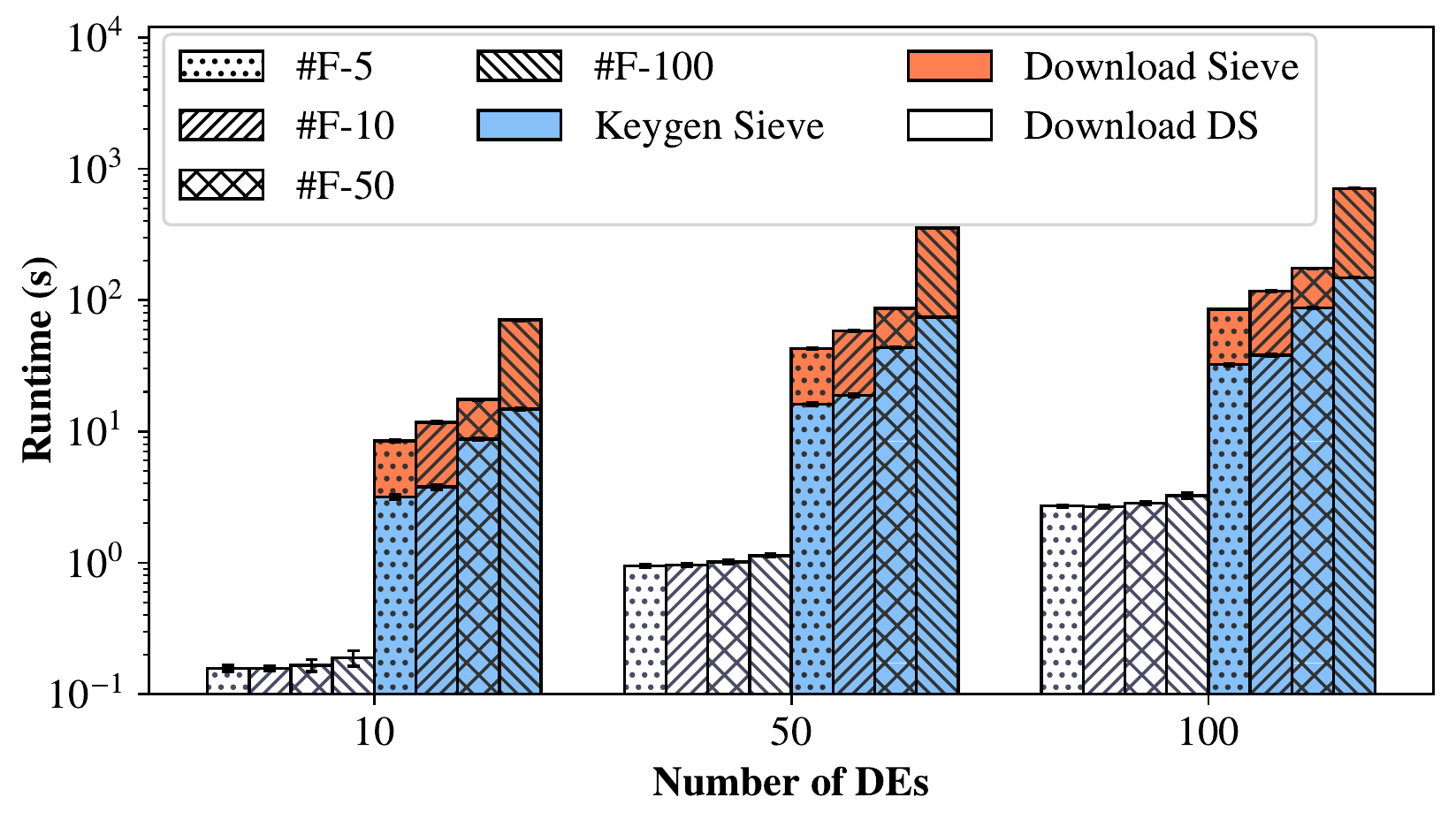}
  \label{fig:ds_sieve_download}
  }
  \centering
  \subfloat{
  \includegraphics[width=0.33\linewidth]{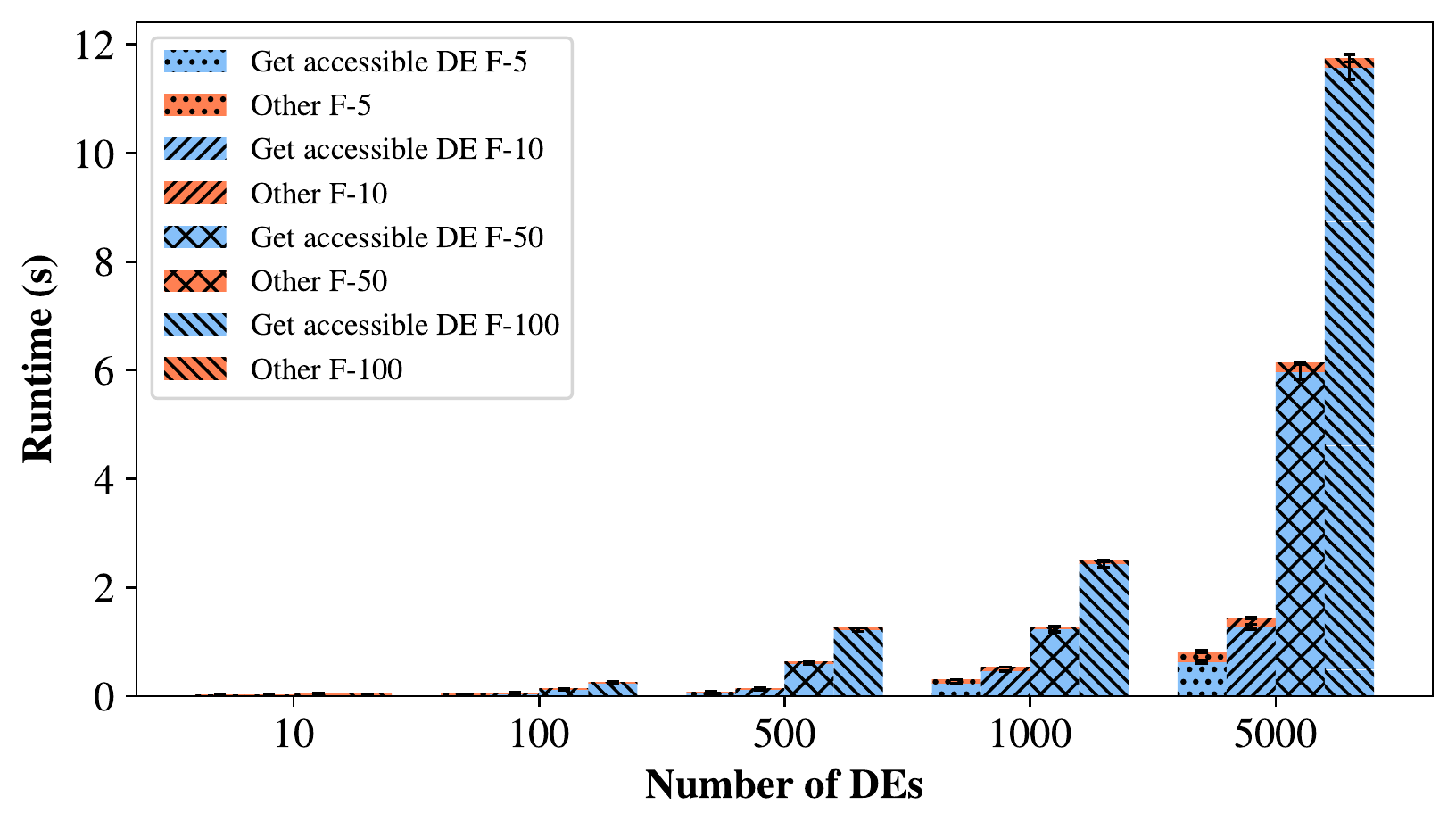}
  \label{fig:ds_user_overhead}
  }
  \centering
\caption{Left (Center): Upload (Download) time of Data Station and Sieve. Right: Data Station overhead on data users}
\label{fig:rq2_stability}
\end{figure*}

\mypar{Machine Learning tasks} \update{R1D4}{We consider 3 machine learning tasks:}

% \begin{myitemize}

\noindent$\bullet$ \update{}{Logistic Regression on the Income dataset}~\cite{income}. \update{}{The task is to predict whose salary is $<$\$50K. In the federated learning baselines, each agent has access to a 4.5K-sample even split of the dataset.}

\noindent$\bullet$ \update{}{Computer vision on the CIFAR-10 dataset}~\cite{krizhevsky2009learning}. \update{}{Each agent has access to 6,250 samples.} 

\noindent$\bullet$ \update{}{Computer vision on the COVIDx CXR-3 dataset}~\cite{covid-net}. \update{}{Each agent has access to 3810 chest-ray images. The task is to predict whether the patient has COVID-19. The total size of the dataset is 14GB and the model used is neural network. }

% \end{myitemize}

\mypar{Experimental Setup} \update{R1D5}{We use n2d-highmem-8 instances (8 virtual CPUs and 64GB of RAM) from Google Compute Engine. In particular, we use 8+1 (clients+server) such nodes for \textsf{Flower}; in the case of \textsf{OpenFL}, we run all clients and server in a single node because the framework does not work well in a distributed setting. We run \textsf{Base} and \textsf{DS} in a single node with the same specs as the other nodes, but enabling SEV when running \textsf{DS} to support \emph{near-zero-trust} mode. We use the same logic for pre-processing, training, inference, and evaluation across baselines, with one exception. OpenFL on income is implemented using a Keras model (instead of sklearn which we use in the others) becasue of the limited support of the framework for other libraries.}

\mypar{Metrics and Baselines} We evaluate the time it takes to train the model and the maximum accuracy achieved in the four baselines. 

% baseline. We measure three baselines, Data Station, federated learning, and a plain model training without encryption and assuming all data is centralized.

% \begin{figure}
%     \centering
%     \includegraphics[width=\columnwidth]{img/ml/income_8.pdf}
%     \caption{Accuracy vs Runtime of Data Station (DS-LR), Flower (FL-LR), and centralized (Basic-LR) logistic regression models on the Adult Income dataset.}
%     \label{fig:ml_income}
% \end{figure}

\mypar{Results} \F\ref{fig:ml_income}, \F\ref{fig:ml_cifar}, \F\ref{fig:ml_covid} \update{R1D3, R1D4}{show the results of the experiment for the three tasks. The performance of \textsf{Flower} and \textsf{OpenFL} is shown as a convergence line. We fit a line to increase readibility when the underlying convergence behavior is spiky. We use dots that indicate when convergence is achieved for \textsf{Base} and \textsf{DS}. We observe several trends.}

\update{}{First, on the same amount of time, centralized training (\textsf{Base} and \textsf{DS}) achieve higher accuracy than the federated learning baselines. Second, the federated learning baselines take 3x more time (e.g., 11 hours vs 4 hours in} \F\ref{fig:ml_covid}) \update{}{to achieve a similar accuracy than \textsf{DS}. This is true for the more complex deep network models: CIFAR} (\F\ref{fig:ml_cifar}) \update{}{and COVIDx} (\F\ref{fig:ml_covid}). \update{}{In the much simpler, logistic regression model used for the income experiment, the centralized approaches and Flower perform similarly well; OpenFL takes a bit longer to produce results but otherwise achieves a similar accuracy.}

\update{}{Second, both federated learning frameworks perform similarly. OpenFL is less stable and runs out of memory in the COVIDx experiment (the X in} \F\ref{fig:ml_covid}). \update{}{Note that we run the federated learning baselines in their \emph{best-case-scenario}, where each client has a random sample of the training data. When this is not true (common in practice) the efficiency of federated learning reduces.}

\update{R1D3, R1D4}{Third, the runtime overhead of \textsf{DS} with respect to \textsf{Base} is minimal across all datasets, despite running in \emph{near-zero-trust} and maintaining clients' data encrypted end-to-end. However, \textsf{DS} running in \emph{near-zero-trust} cannot run computation on GPUs, so when doing so brings performance benefits, \textsf{DS} leaves those on the table. We demonstrate that in} \F\ref{fig:ml_covid},\update{R1D3, R1D4}{ by including \textsf{Base} running on a GPU and showing the performance difference with \textsf{DS}.}

The results show the advantages in accuracy and runtime of centralizing data and compute, and the low overhead of Data Station compared to off-the-shelf (non-trusted) model training.

\subsubsection*{Qualitative Analysis}

\update{R2D1}{There are various important qualitative differences between Data Station and federated learning:}

\mypar{Compatibility} Federated learning supports only machine learning models that can be merged. Data Station has no such limitation. Similarly, \update{R2D1}{existing applications must be modified and adapted to the federated learning framework. Data Station only requires writing a simple connector to expose the functions.}

\mypar{Security and Leakages} Many federated learning frameworks leak weight updates~\cite{kairouz2021advances}, which can be leveraged to reconstruct part of the data, thus breaking the promised security guarantees. A solution is combining federated learning with differential privacy~\cite{wei2020federated}. This requires further modifications to applications and algorithms, and results in further performance reduction.

\mypar{Performance Differences} Many federated learning algorithms do not guarantee the same performance as centralized implementations. We demonstrated this quantitatively above.

\mypar{Scalability} \update{R2D1}{While (in principle) federated learning scales with the number of data contributors, the current design of Data Station would need to adapt to support multi-machine setting. We do not anticipate severe barriers in achieving that goal.}

\begin{figure*}[]
  \centering
  \subfloat{
  \includegraphics[width=0.33\linewidth,,valign=t]{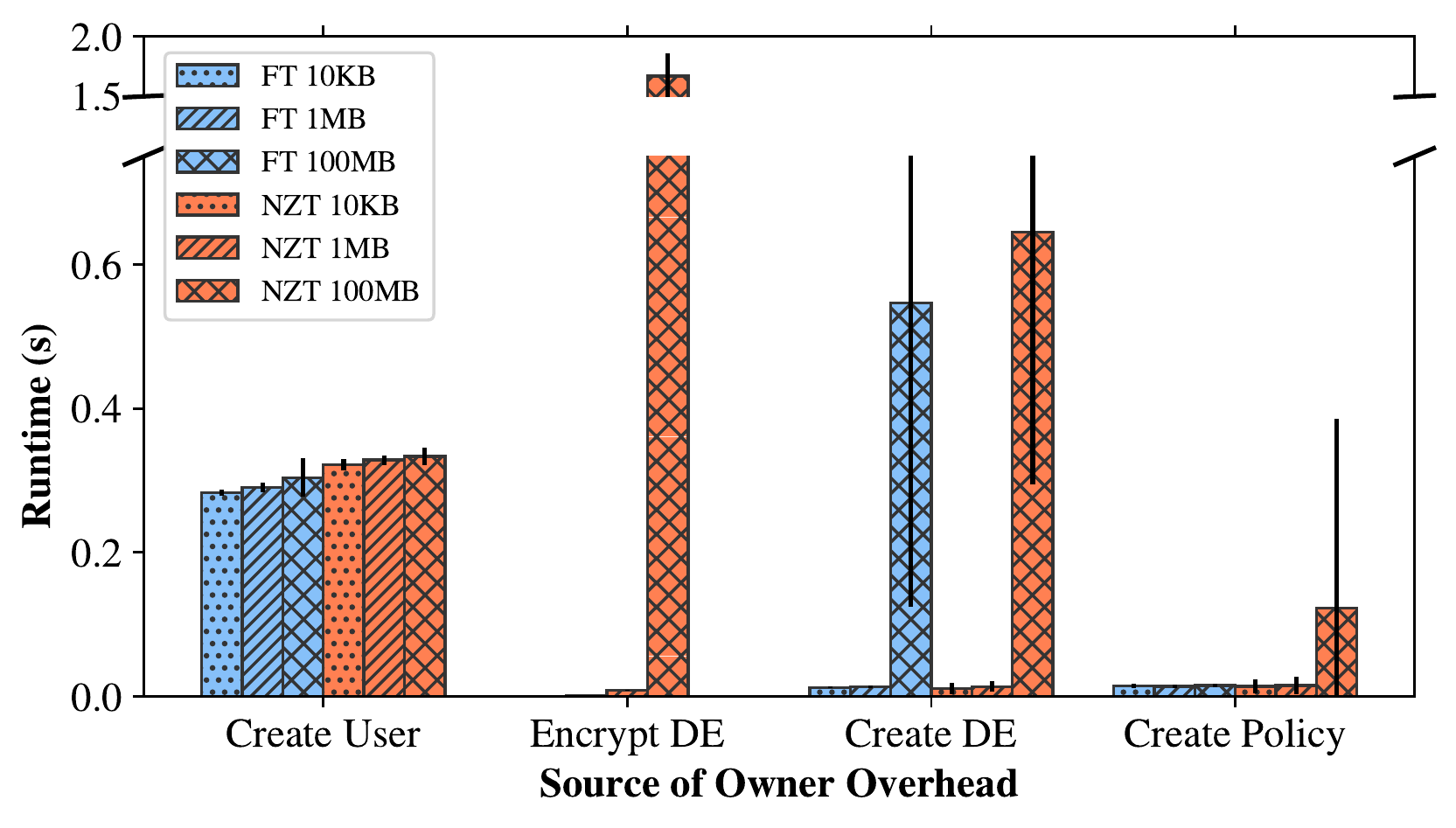}
  \label{fig:ds_owner_overhead}
  }
  \subfloat{
  \includegraphics[width=0.33\linewidth,,valign=t]{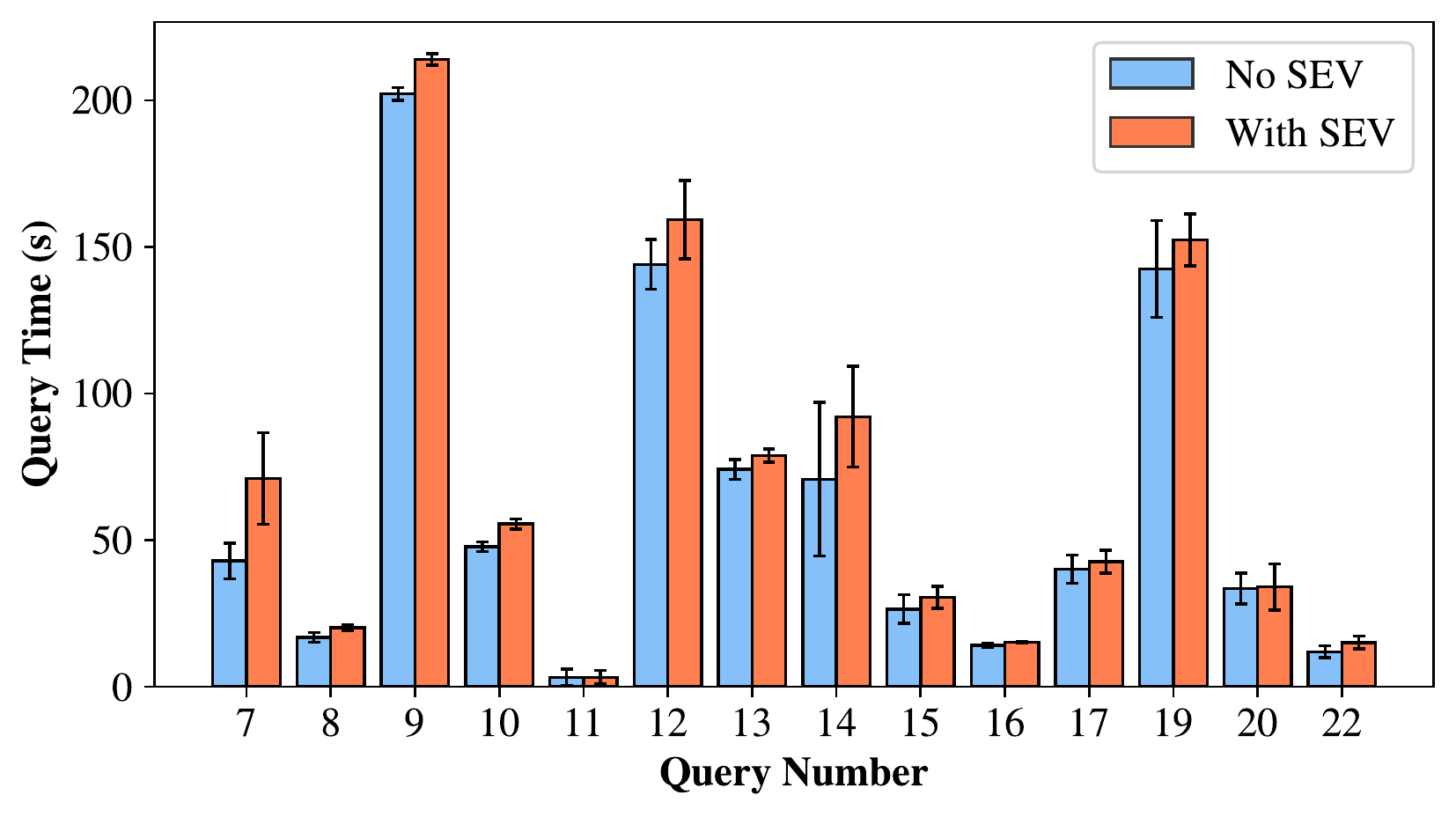}
  \label{fig:tpch_benchmark}
  }
  \centering
  \subfloat{
  \includegraphics[width=0.33\linewidth, ,valign=t]{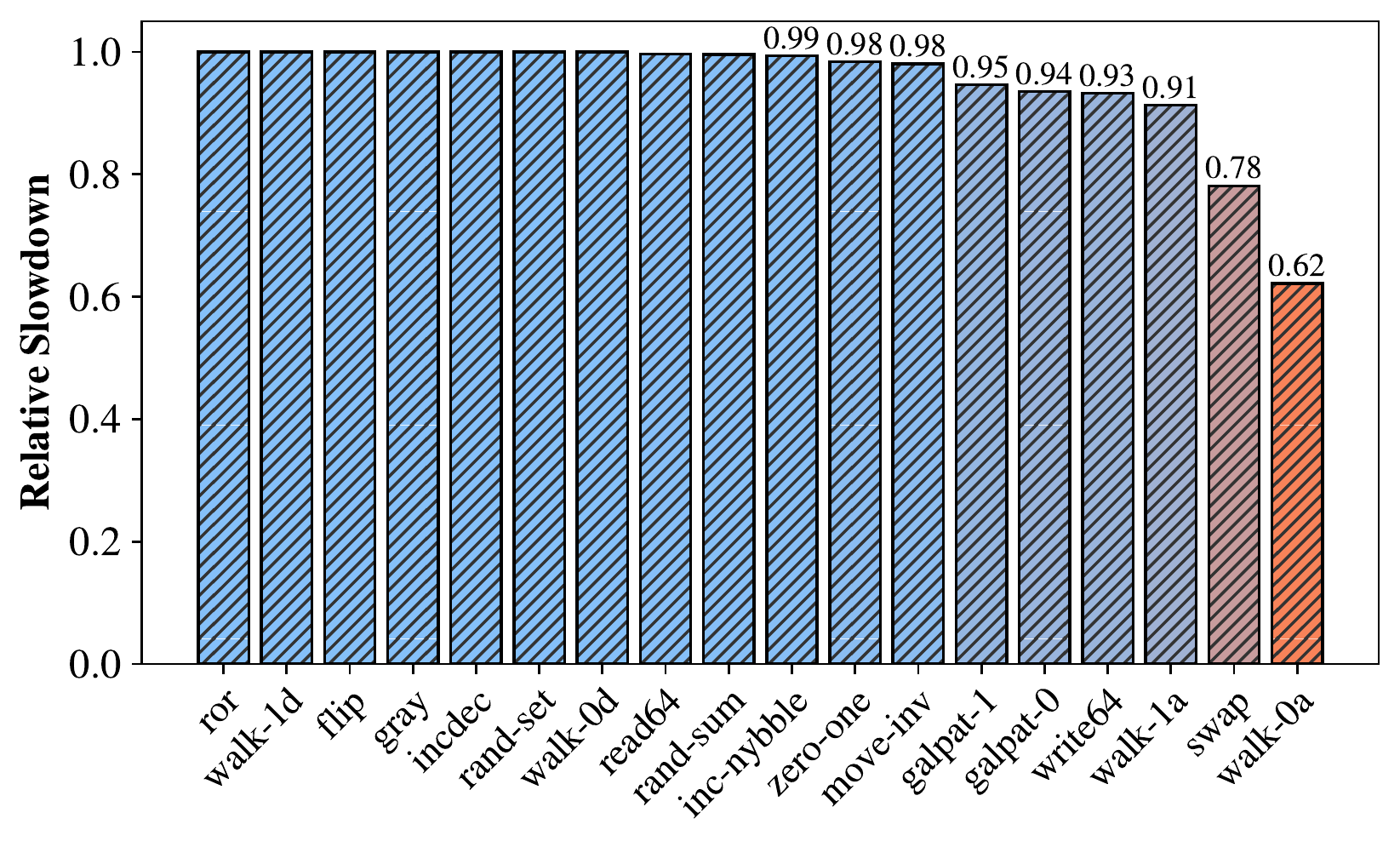}
  \label{fig:stressng_benchmark}
  }
  \centering
\caption{Left: Overhead for Data Owners. Center: SEV Overhead on TPC-H Benchmark. Right: stress-ng Benchmark.}
\label{fig:rq2_stability}
\end{figure*}

\subsection{Secure Data Sharing}
\label{subsec:secure_data_sharing}

\update{R2D3}{In this scenario, owners store data in a third-party server and selectively let other agents access this data by writing policies. In this scenario, we execute Data Station in \emph{near-zero-trust} mode, and so it is subject to overheads from the enclave and the encryption protocol. We want to understand whether these overheads lead to an impractically slow platform. To answer that question, we measure the runtime and compare it with Sieve's}~\cite{wang2016sieve}.

\mypar{A Primer on Sieve} Sieve enables cryptographically secure data sharing via an untrusted intermediary. It uses symmetric keys for encrypting files, like Data Station. It encrypts metadata for each file with attribute-based-encryption (ABE) \cite{goyal2006attribute} and this allows it to generate decryption keys that only work with selected attributes. Finally, it relies on homomorphic encryption \cite{gentry2009fully} for revoking access to datasets. Unlike Data Station, Sieve is exclusively designed for file sharing. \update{R1D3}{Sieve is the only approach we identified that simultaneously addresses the secure data sharing problem, is open source, and uses cryptographic techniques to build trust with data owners. The role of Sieve in our evaluation is to provide a reference performance to facilitate interpreting the performance of Data Station when running in \emph{near-zero-trust} mode.}

\mypar{Experimental Setup} Sieve can generate decryption keys that only decrypt data previously encrypted with specific attributes. We use one Sieve attribute per registered function inside Data Station. Assume there are $n$ DEs and $m$ functions. Sieve encrypts each dataset with a symmetric key and the metadata (the functions) with ABE. This lets a user who wants to invoke a function $f \in m$ download a dataset in Sieve. Data Station runs in \emph{near-zero-trust} mode and implements a \textsf{download} application. \textsf{download} takes a DE as input parameter and, if permitted by the gatekeeper, sends the DE to the user. This replicates the Sieve setup with the same functionality, thus letting us compare both approaches.

% \mypar{Experimental Setup} We run Data Station and Sieve in a n2d-highmem-8 instance of Google Compute Engine, with 8 virtual CPUs and 64GB of RAM. We run Data Station on an image with SEV enabled and Sieve without SEV, thus avoiding any overheads that may be introduced by the secure enclave.

\mypar{Metrics} We evaluate the end-to-end performance of data sharing by measuring the time for the data owner to upload datasets and the time for the data user to download datasets. In Data Station, we create $n*m$ policies to indicate that a user can invoke any function on any DE. We use files each consisting of 10KB random byte strings. We show average runtime over 20 runs.

% run every experiment 20 times and show average and error bars. 

% To simulate the setup of Sieve, assuming there are $n$ DEs and $m$ functions that need to be shared and each DE has $m$ attributes in Sieve, $m$ functions will be registered inside Data Station and the operations they perform are exactly the same -- downloading the content of all DEs. 

% In Data Station, the data owner will create $nm$ policies for a data user, so that the user can access all $n$ DEs using any one of the $m$ functions. Note that $nm$ is the maximum number policies one can create in Data Station given $n$ DEs and $m$ function, thus we are effectively comparing the performance of Data Station in the worst case scenario.

% \begin{figure}
%     \centering
%     \includegraphics[width=\columnwidth]{img/file_sharing/upload.pdf}
%     \caption{Upload time of Data Station and Sieve}
%     \label{fig:ds_sieve_upload}
% \end{figure}

\mypar{Upload Results} \F\ref{fig:ds_sieve_upload} \update{R1D10}{shows the results for uploading data with Data Station and Sieve when changing the number of DEs (x axis) and the number of registered functions (different bar styles).} First, when the number of registered functions (\#F-$m$) is small, Data Station is an order of magnitude faster than Sieve. When the number of functions increases to 50, Data Station is 80\% faster than Sieve, and at 100 functions Data Station is still 25\% faster. Second, the scaling behavior in both Data Station and Sieve is similar and depends on the number of DEs and functions. However, what is crucially important is that, in Data Station, we are measuring the \emph{worst case scenario} where every DE can be downloaded by any function. Note that, when this is not the case, Data Station's overhead will reduce (with the number of DEs and actual functions allowed), while in the case of Sieve it will remain constant because it still needs to encrypt a value for the attribute. Finally, in Data Station the dominating cost is creating policies that requires writing to the EWAL as it operates in \emph{near-zero-trust}. In Sieve, the symmetric encryption of the file varies only with the number of DEs uploaded. The most significant Sieve overhead is the ABE encryption time, accounting for over 95\% of the upload time among all DEs and attributes. Data Station outperforms Sieve's even in the worst-case scenario of allowing every function to execute on every DE.

% At a low number of functions, Data Station significantly outperforms Sieve, and at the worst case of 100 functions, Data Station' upload rate is still comparable and slightly higher.

% simulates an agent uploading a set of DEs to the intermediary. For each trial, we vary the number of attributes and DEs (10KB each). For Data Station, we show both the time to upload DEs and the time to create policies. For Sieve, we separate the ABE time from the rest of the uploading process to demonstrate any bottlenecks.

% As shown in \F\ref{fig:ds_sieve_upload}, the runtime of uploading DEs and creating policies combined in Data Station is faster than the total upload time in Sieve across all numbers of attributes and DEs. For Data Station, creating policies takes the majority of time since we are creating the maximum number of policies.

% Wordle Colorblind colors :D
% \begin{figure}
%     \centering
%     \includegraphics[width=\columnwidth]{img/file_sharing/download.pdf}
%     \caption{Download time of Data Station and Sieve}
%     \label{fig:ds_sieve_download}
% \end{figure}

\mypar{Download Results} \F\ref{fig:ds_sieve_download} \update{R1D10}{shows the results for downloading data when changing the number of DEs (x axis) and the number of registered functions (different bar styles).} After a user calls \textsf{download} and the \textsf{Gatekeeper} checks they have access the function reads the encrypted DE into memory, decrypts it with the owner's symmetric key, re-encrypts it with the user's symmetric key and sends it to the user. Sieve's higher runtime stems from the use of ABE. Sieve's owner has to generate a decryption key (Keygen) via ABE for the necessary attributes, shown in the bottom bar of \F\ref{fig:ds_sieve_download}. The generated key is sent to the user, who uses it to decrypt the data; this decryption accounts for over half of the time spent downloading. Although Sieve's design allows the key generation to execute only once for a user to gain access to the data, the decryption and download costs must still occur. Data Station thus provides two benefits over Sieve: i) owner needs no interaction with the user downloads; ii) runtime is between 1 and 2 orders of magnitude lower.

% Once the agent has uploaded the set of DEs, afterwards another agent would download them. For each set of DEs, the uploaded data was subsequently downloaded while timed, shown in \F\ref{fig:ds_sieve_download}; we separate the key generation time, necessary for matching the attributes to decrypt in Sieve, from the rest of the download time. 

% As shown in \F\ref{fig:ds_sieve_download}, the download time in Data Station is almost unnoticeable compared to Sieve. Although downloading DEs in Data Station has to go through the \textsf{Interceptor}  to decrypt the data, the process is still orders of magnitude faster than both the key generation time and actual download time (which also involves decryption) in Sieve.

% \notechz{Do we want any analysis of just Sieve itself, or should we only include comparisons to DS?}

\subsubsection*{Qualitative Analysis}

We complement the quantitative results with important qualitative differences between approaches.

\mypar{Trust Model} Unlike Data Station, Sieve does not hide file metadata (\ie policies) from the infrastructure provider. Thus, a curious provider will learn the functions available in the platform and who has access to them, even though they cannot read the raw data. Data Station protects this information by keeping the database, EWAL, and audit log encrypted at all times. A curious provider may learn agent IDs but cannot associate them with users.

\mypar{Delegated Computation} Sieve does not support delegated computation, it acts as a storage provider. The implication is that once a user is given access to a DE, the owner in Sieve must connect directly to the user. This is unlike Data Station, where we can provide a function to perform this delegated computation, in this case the simple \textsf{download} function, but in general, more complex functions as we saw in the machine learning application above.

\mypar{Revoking Access to Data} In Data Station, revoking access to data translates to modifying a policy, which is a quick operation, and can be done selectively for a given user. In contrast, in Sieve this involves changing the attributes for the previously uploaded data and then re-encrypting the file on the storage provider. Because the file cannot be decrypted first, Sieve uses a clever technique to re-encrypt it without decryption using homomorphic encryption. This has several implications. First, the newly generated key must be re-sent to those users who still have access to the file. Second, the revocation process itself is much slower than in Data Station because of the aforementioned techniques.

Sieve is a well designed and engineered system. We believe the results are representative of solutions fully based on cryptographic techniques and thus validate Data Station's design. 

% Revoking access to a DE for a single agent in Sieve involves changing the attributes associated with that data, thus invalidating any keys associated with that set of attributes. The consequences are that any agent still permitted to access the data must be given a new key, and if each key is unique, a new one must be generated for all of them, making time for redistributing access linear with the number of agents. \notechz{DS in contrast is roughly constant time, since it only requires a database update to revoke a policy.}

\subsection{Data Station Performance Analysis}

\update{R1W6}{The previous two experiments demonstrate that the overhead Data Station introduces is small compared to application time. Here, we use a \textsf{noop} function to understand Data Station's \emph{fixed} overheads for users and owners. We conclude by measuring the overhead AMD SEV introduces in Section}~\ref{eval:sev}.

% We implement a \textsf{noop} application for Data Station that performs the noop operation. We then measure the overhead for data users when calling this function (Section~\ref{eval:useroverhead}) and the overhead for owners when setting up the Data Station in Section~\ref{eval:owneroverhead}. We conclude by measuring the overheads of hardware enclaves in Section~\ref{eval:sev}.

% \steven{It is important to note that as we are performing the noop operation, the overhead for data users does not depend on the attributes of the DEs, because Data Station is not accessing the DEs (and consequently, is not performing any decryptions). As a result, the overhead for \emph{full-trust} mode and \emph{near-zero-trust} mode is largely the same. So we only present the overhead for \emph{full-trust} mode for data users. The single most important attribute of DEs that affects Data Station's overhead is the size of the DEs, as it affects the decryption time. Its effect has been investigated in prevoius experiments already. On the other hand, for data owners, we measure the overhead in both \emph{full-trust} mode and \emph{near-zero-trust} mode.}

% \begin{figure}[h]
% \begin{mdframed}[backgroundcolor=yellow!50,linecolor=yellow!50]
%     \centering
%     \includegraphics[width=\columnwidth]{img/microbenchmarks/user_overhead.pdf}
%     \caption{Overhead for Data Users}
%     \label{fig:ds_user_overhead}
% \end{mdframed}
% \end{figure}

\subsubsection{Overhead for Data Users}
\label{eval:useroverhead}

\update{R1W6, R1D6}{After a user invokes a function, the potential sources of overhead are: i) obtain user's ID from the database; ii) obtain accessible DEs from policy broker; iii) initialize execution environment; iv) collect list of actually accessed DEs after function invocation; and v) log all activity in the audit log.}

\mypar{Overhead Factors} \update{R1W6, R1D6, R1D7}{In the \texttt{noop} operation, no DE is ever accessed so no DE characteristics (e.g., data type) affect the overhead. The DE size has an effect when they need to be decrypted (when operating in near-zero-trust mode), but we already studied such an effect in Section}~\ref{subsec:secure_data_sharing}. \update{}{Here, we concentrate instead on the 5 steps above. Of those, 'obtaining the list of accessible DEs' dominates the overhead, and there are two factors that affect such overhead: the number of registered DEs and the number of functions registered. Consequently, we show results varying these two factors.}

\mypar{Experimental Setup} \update{R1W6, R1D6}{When varying the number of functions registered, we create a policy for each pair of DE and function, and store these policies in the database. We report the average end to end runtime over five runs. Neither DE size nor type affect overhead; we use 1MB DEs representing text files.}

\mypar{Results} \F\ref{fig:ds_user_overhead} \update{R1W6, R1D6}{shows the runtime per function invocation when changing the number of DEs (x axis) and for different numbers of functions. We report the runtime for obtaining the list of accessible DEs, and we group the other sources of overhead together and refer to them as \textsf{Other}. The overhead for getting the accessible DEs scales linearly with the number of policies. With 5000 DEs and 100 functions (i.e., half a million policies) the overhead is only 12 seconds. Contrast that with the multiple minute (and hours) overhead of training a machine learning application such as that in the previous section, which has only a few registered functions. We conclude that the overhead for data users is negligible for applications with runtimes larger than a minute.}

% \steven{\F\ref{fig:ds_user_overhead} shows runtime per function invocation. The points on the x-axis correspond to Data Station with 10, 100, 500, 1000, and 5000 DEs. We measure overhead for function dependency graphs with 5, 10, 50, and 100 nodes, respectively. The blue bars represent the overhead of calling the \textsf{Policy Broker} to obtain the list of accessible DEs, and the orange bars represent the total overhead of performing the other four steps. Note that the overhead of performing the other four steps remain negligible, comparing to getting the accessible DEs. The overhead for getting the accessible DEs scales linearly with the number of policies as expected. For Data Station with 5000 DEs, and an application with a dependency graph with 100 nodes (corresponding to 100 functions), the overhead is 12 seconds\notera{ create new par to interpret these results. first, contrast this with the apps above. second, give an example with the sharing consortia with 8 individuals and show what the overhead is in that case.}. Overall, these overhead is negligible: note that the applications we presented above have fewer than 5 nodes in their function dependency graph.}

\subsubsection{Overhead for Data Owners}
\label{eval:owneroverhead}

\update{R1W6, R1D6}{The sources of overhead for data owners are: i) registering with the platform; ii) (only in near-zero-trust mode) encrypting the DEs to upload; iii) uploading DEs; and iv) creating policies. Unlike before, the size of DE matters because of sources ii) and iii), so we vary the DE size in this experiment.} 

% \notezz{if the experimental setup differs from 6.3.1 (here it mentioned the DE size is varied), then maybe we should also include a small paragraph about experimental setup here}

% \steven{Running an application in the Data Station requires registering the agents, uploading the DEs and policies. In \emph{near-zero-trust} mode, it also requires encrypting the uploaded DEs. These operations form the overhead for data owners. Thus, we measure the overhead for data owners in terms of these three operations, in both \emph{full-trust} mode, and \emph{near-zero-trust} mode.}

% \begin{figure}
%     \centering
%     \includegraphics[width=\columnwidth]{img/tpch.pdf}
%     \caption{SEV Overhead on TPC-H Benchmarks}
%     \label{fig:tpch_benchmark}
% \end{figure}

% \begin{figure}[h]
% \begin{mdframed}[backgroundcolor=yellow!50,linecolor=yellow!50]
%     \centering
%     \includegraphics[width=\columnwidth]{img/microbenchmarks/owner_overhead.pdf}
%     \caption{Overhead for Data Users}
%     \label{fig:ds_owner_overhead}
% \end{mdframed}
% \end{figure}

\mypar{Results} Figure~\ref{fig:ds_owner_overhead} \update{R1W6, R1D6}{shows the sources of overhead in the x axis. We show results for DEs of different sizes to expose the overheads introduced by encryption (\textsf{Encrypt DE}) and uploading the DE (\textsf{Create DE}). We report average over 100 runs. The overhead of \textsf{Create User}, \textsf{Create DE} and \textsf{Create Policy} is always slightly higher in \emph{near-zero-trust} mode due to the use of SEV and writing to the EWAL, but remain sub-second. The largest source of overhead is encrypting DEs (only in \emph{near-zero-trust} mode), which depends on DE size.}

% \steven{The results are shown in Figure~\ref{fig:ds_owner_overhead}. The points on the x-axis correspond to the overhead of creating users, encrypting DEs (only in \emph{near-zero-trust} mode), creating DEs, and creating policies. We measure the overhead for synthetic dataset of size 10KB, 1MB, and 100MB, respectively. We vary the size of the DEs because it affects the time to do encryption, as well as creating DEs (as Data Station need to copy the content of the DEs). For each bar, we performed the operation 200 times, and took the average of those. We observe that the time of encrypting DEs and creating DEs becomes more dominant, as the size of DE increases. For all these operations, there is a slight additional overhead of \emph{near-zero-trust} over \emph{full-trust} mode, caused by using SEV and writing to EWAL. Overall, these overheads remain small.}

\subsubsection{AMD SEV's Performance Overhead}
\label{eval:sev}

% \begin{figure}
%     \centering
%     \includegraphics[width=\columnwidth]{img/tpch.pdf}
%     \caption{SEV Overhead on TPC-H Benchmarks}
%     \label{fig:tpch_benchmark}
% \end{figure}

We measure the overhead introduced by AMD's SEV \update{R1D8}{using the TPC-H benchmark, as a representative data-intensive workload familiar to the reader, and stress-ng}~\cite{stress-ng}, \update{R1D8}{a tool to stress computer systems and used in previous work to understand enclave's performance}~\cite{gottel2018security}. We run TPC-H on DuckDB~\cite{raasveldt2019duckdb} over a 100GB database with and without SEV. \F\ref{fig:tpch_benchmark} shows the average runtime for both baselines after 10 runs of each query. The results show that SEV indeed introduces overhead; \update{R1D8}{most noticeable in query 7 that writes a large amount of disk-resident data into (encrypted) memory}. However, the overhead is modest given the security guarantees gained. The results cement the advantages of modern secure hardware enclaves and the opportunities that they open.

\update{R1D8}{stress-ng }\cite{stress-ng} \update{}{uses \emph{stressors} to understand system performance.} \F\ref{fig:stressng_benchmark} \update{}{shows the relative slowdown of a stressor with SEV active, measuring the number of operations completed per unit of time} ($\text{slowdown} =  \frac{\text{SEV ops done}}{\text{no SEV ops done}}$). \update{}{The stressors that introduce the largest overhead (e.g., walk-0a, swap, walk-1a) correspond to those performing random-access, memory-intensive tasks. The \textsf{walk-Xa} stressors force reads from physical memory, which explains the larger overhead, and \textsf{swap} exchanges the contents of two different memory locations.}

\section{Related Work}
\label{sec:relatedwork}

Data Station is the first data escrow system that concentrates in offering delegated, trustworthy, and auditable computation. It builds on many existing lines of research that we explain below.

\myparnoperiod{Hippocratic Databases}~\cite{agrawal2002hippocratic}. \update{R3}{Data Station is related to the Hippocratic databases vision. There are important similarities and differences. Crucially, in both papers access control is defined around a \emph{purpose}, which is specified in Data Station with a policy referencing a function. In this way, both vision and system chase a contextual integrity}~\cite{nissenbaum2004privacy} \update{}{view of privacy more so than one based on access control}~\cite{westin1968privacy}. \update{}{Another similarity is the need for auditable computation, via the \textsf{audit log} in Data Station and the concept of audit trails in Hippocratic databases. The differences are also important. Hippocratic databases are envisioned as a database system that takes care of privacy. Data Station is a data escrow system for implementing other applications, including RDBMS but also ML and other analytics-based functionality. As Data Station decouples computation and data, a crucial element is the need to build trust with users and owners alike.} 

% \steven{I think we can add one more sentence to strengthen the difference, but I don't know how.}

\myparnoperiod{Data Enclaves} store restricted-use data, \eg data subject to privacy and regulation constraints~\cite{foster2018research}. The enclave ensures that the hosted data is protected while permitting users to execute certain pre-determined computations, often with review prior to data release~\cite{lane2008data,zeng2014cloud}. They are commonly found in research organizations, where they are built with the intention of facilitating data-driven research, e.g., ICPSR in the social sciences~\cite{icpsrenclave}, and NORC~\cite{norcenclave}. These enclaves are the result of long sustained efforts. Data Station is geared towards easing the creation of data enclaves, including to share data among organizations.

\mypar{Multi-Party Computation, Homomorphic Encryption, Federated Learning, and Differential Privacy} There is a growing class of systems designed to permit collaborative analytics on restricted-use datasets. Shrinkwrap~\cite{bater2018shrinkwrap} and Saqe~\cite{bater2020saqe} permit the execution of SQL queries over data from multiple organizations, while ensuring differentially private results that do not disclose the identity of any participant. Conclave~\cite{volgushev2019conclave}, Cerebro~\cite{zheng2021cerebro}, and Secrecy~\cite{liagouris2021secrecy} rely on multi-party computation to achieve a similar goal. Alternative techniques such as homomorphic encryption permit run computation directly on encrypted data. These technologies concentrate on providing trustworthy data processing. Unlike existing solutions and technologies, Data Station is designed to tackle the three requirements presented above, which no other solution achieves.

% Other approaches rely on multi-party computation to achieve a similar goal, such as Conclave~\cite{volgushev2019conclave}, Cerebro~\cite{zheng2021cerebro}, and Secrecy~\cite{liagouris2021secrecy}. Other techniques, such as homomorphic encryption, permit certain computation on encrypted data, which may be used to build trust with data owners. These technologies and systems concentrate on securing data processing even when it takes place in third-party infrastructure, \ie enabling trustworthy computation. Unlike existing solutions and technologies, Data Station is designed to tackle the three requirements presented above, which no other solution achieves.

\myparnoperiod{Confidential Computing} initiatives have been promoted by cloud vendors to build trust with customers who store their data in the cloud. Azure Always Encrypted~\cite{antonopoulos2020azure} lets users run computation on data they own on infrastructure they do not own (\ie the cloud). Similarly, research approaches such as Haven~\cite{haven}, SCONE~\cite{arnautov2016scone}, Keystone~\cite{lee2020keystone}, Ryoan~\cite{hunt2018ryoan}, VC3~\cite{schuster2015vc3}, protect data processing when the data is hosted in the cloud. Many of them focus on protecting databases~\cite{fuhry2021encdbdb, zhu2021full, sun2021building, zheng2017opaque}. Some recent work has proposed infrastructure to facilitate running SQL queries across data from multiple parties~\cite{dave2020oblivious}. All these solutions leverage secure hardware enclaves, like Data Station. But unlike Data Station, none of them enable delegated, trustworthy, and auditable computation on data owned by multiple parties.

% \smallskip

\mypar{Auditability} Many have noted the importance of building audit logs, in IoT environments~\cite{panwar2021iot}, using blockchain technology~\cite{russinovich2019ccf, shamis2021pac, maiyya2019database} to make the log tamper-proof, and even building databases on top of that abstraction~\cite{el2019blockchaindb}. Similarly, there is work proposing ways of reasoning about access policies in the context of databases~\cite{amsterdamer2020towards}. The techniques used in these approaches are complementary to the ones we use in the audit log in Data Station. Unlike these approaches, Data Station leverages the centralization of data and compute to simplify the problem.

% \subsection{Decentralized architectures}

% - Bringing Decentralized Search to Decentralized Services (find other relevant papers by Sebastian Angel) \cite{li2021bringing}

% - “Privacy Preserving Vertical Federated Learning for Tree-based Models”.\cite{wu2020privacy}

% \subsection{Other relevant work to classify}

% - IoT Notary: Attestable Sensor Data Capture in IoT Environments

% - carsten blockchaindb \cite{el2019blockchaindb}

% \notera{the microsoft system from miguel castro and co}

\section{Conclusions}
\label{sec:conclusions}

The increasing number of scenarios where organizations benefit from pooling and sharing data calls for technical solutions to ease the task of forming data-sharing consortia. We have presented Data Station, a data escrow system that implements delegated, trustworthy, and auditable computation with the aim of facilitating owners and users to share and benefit from each others' data. Data Station supports unmodified applications and thus a wide range of applications. We presented mechanisms to generate trust from owners and users based on the use of secure hardware enclaves and cryptographic protocols. \update{R1W1, R1D1}{The evaluation results demonstrate the feasibility of the approach when compared to strong baselines for other applications, including machine learning training consortia and secure data sharing. We study the overheads and demonstrate they are negligible compared to application runtime. Beyond the quantitative differences, we highlight important qualitative advantages of the data escrow design.}

% \input{sections/extra}
% \input{sections/auction}

% \begin{acks}
%  This work was supported by the NSF Convergence Accelerator (Award Number \#2040718). We thank Frank Wang for helping us set up Sieve and Christian Goettel for helping us benchmark SEV. We also thank the Chameleon cloud~\cite{keahey2020lessons} for providing computational resources that helped with the evaluation and testing of our system, and the anonymous reviewers for providing valuable feedback that helped us improve the paper.
% \end{acks}

\bibliographystyle{ACM-Reference-Format}
\bibliography{main}

\end{document}